\pdfoutput=1

\documentclass[11pt]{article}

\usepackage[final]{acl}

\usepackage{times}
\usepackage{latexsym}

\usepackage[T1]{fontenc}

\usepackage[utf8]{inputenc}

\usepackage{microtype}

\usepackage{inconsolata}

\usepackage{graphicx}

\usepackage{amsmath,amssymb,amsthm}
\usepackage{algorithm,algorithmic}
\usepackage{booktabs}
\usepackage{hyperref}
\usepackage{enumitem}
\usepackage{tikz}
\usetikzlibrary{positioning,arrows.meta,shapes,fit,backgrounds,calc}
\usepackage{pgfplots}
\pgfplotsset{compat=1.18}

\newtheorem{theorem}{Theorem}
\newtheorem{proposition}{Proposition}
\newtheorem{corollary}{Corollary}
\newtheorem{definition}{Definition}
\newtheorem{lemma}{Lemma}
\newtheorem{remark}{Remark}
\newcommand{\cmark}{$\checkmark$}
\newcommand{\xmark}{$\times$}

\title{Option-GRPO: Hierarchical RL Against AI Slop in Code Generation for Unseen SDKs}

\author{Prashank Kadam \\
  Cerby Inc., USA \\
  \texttt{prashank@cerby.com}}

\begin{document}

\maketitle

\begin{abstract}
Large language models increasingly produce \emph{AI slop} in generated code: verbose, redundant, non-idiomatic implementations that pass functional checks but bloat codebases. The problem compounds on SDKs released after the model's training cutoff, where the model has neither idiomatic priors nor compositional understanding. We present \textbf{Option-GRPO}, a hierarchical reinforcement learning framework that adapts a code LLM to an unseen SDK from its source code alone, without human-written usage examples or test cases, and that explicitly optimizes for both functional correctness and code elegance. Our key insight is that SDK source code contains compositional structure that defines a hierarchy of reusable code patterns (\emph{code options}) which simultaneously serve as the building blocks of a curriculum and the action space for efficient policy optimization. We show that code options induce a valid semi-Markov decision process, give conditional bounds under which shorter option description length implies more concise and, in expectation, more API-efficient programs, and analyze under stated assumptions how the option hierarchy improves credit assignment by a factor of $\sqrt{T/K}$. Using Group Relative Policy Optimization (GRPO) with execution-based rewards at each curriculum level, Option-GRPO discovers SDK idioms, composes them into patterns, and assembles patterns into complete programs. Experiments on 12 unseen SDKs with a 32B-parameter base model demonstrate $61.4\%$ pass@1, outperforming flat GRPO by $8.7$ points, ExploraCoder by $11.2$ points, and ReCode by $6.3$ points. Option-GRPO produces solutions that use $34.7\%$ fewer API calls and are $27\%$ shorter than flat-GRPO outputs while maintaining higher correctness, directly attacking the slop problem.
\end{abstract}

\section{Introduction}

The rapid proliferation of software development kits creates a fundamental challenge for code generation: the libraries developers most need help with are precisely those released after model training cutoffs, where LLM performance degrades substantially~\cite{xia2023automated, zan2022large}; recent benchmarks confirm that performance also degrades on evolving, version-specific library APIs~\cite{misra2025gitchameleon, liu2024codeupdatearena, kuhar2025libevolutioneval}. A second, increasingly visible problem compounds the first: even on familiar libraries, modern code LLMs generate what practitioners have begun to call \emph{AI slop}~\cite{slop2024} (Merriam-Webster's 2025 Word of the Year), output that is fluent line by line yet bloated, redundant, and non-idiomatic when read as a whole. Industry studies of 211~million lines of production code report that duplicated 5+ line blocks grew roughly eightfold between 2020 and 2024, while the share of refactoring edits dropped by 60\%~\cite{liu2025aislop}.

Existing retrieval-augmented and exploration-based approaches address the first problem partially. DocPrompting~\cite{zhou2023docprompting} retrieves documentation; Gorilla~\cite{patil2024gorilla} fine-tunes for API-call prediction with retrieval awareness; ExploraCoder~\cite{hu2024exploracoder} chains multi-step exploration over unseen APIs without training; and ReCode~\cite{wu2026recode} applies rule-based RL to update an LLM's API knowledge after library version changes. None of these methods explicitly optimize for code quality beyond functional correctness, and none discover compositional idioms from SDK structure that would allow the model to write \emph{elegant} solutions in addition to \emph{working} ones.

We observe that SDK code has an inherent hierarchical structure that mirrors the options framework in hierarchical reinforcement learning~\cite{sutton1999between}: API calls are atomic actions, commonly composed call sequences form reusable patterns analogous to \emph{options} (temporally extended actions in HRL), and complete programs are compositions of these patterns. This hierarchy is encoded in the SDK source itself: type relationships and parameter-flow patterns dictate which functions compose, and frequently composable subgraphs correspond to natural programming idioms.

We present Option-GRPO, a framework that formalizes this connection: each stage of a learning curriculum produces reusable code options that become the action space for the next stage. At each level, Group Relative Policy Optimization (GRPO)~\cite{shao2024deepseekmath} with execution-based rewards optimizes the policy without requiring a critic network. The group-relative comparison within GRPO, combined with the option-level action space, provides natural credit assignment: determining which option in a composed sequence caused success or failure becomes tractable when comparing groups of option sequences rather than groups of token sequences.

Experiments across $12$ unseen SDKs demonstrate that Option-GRPO achieves $61.4\%$ pass@1 with a Qwen3-32B base, improving over flat GRPO ($52.7\%$), ExploraCoder ($50.2\%$), and ReCode ($55.1\%$). More notably, generated code is $27\%$ shorter and uses $34.7\%$ fewer API calls, while maintaining higher functional correctness. Analysis shows that options provide sufficient credit assignment for GRPO to converge stably on tasks where flat GRPO fails due to reward sparsity, and the resulting policy reuses idiomatic SDK patterns in $74\%$ of generated tokens, against $41\%$ for flat GRPO.

\paragraph{Contributions.}
\begin{enumerate}[leftmargin=*]
    \item \textbf{Option-GRPO}, a hierarchical RL framework for example-free SDK mastery that formalizes curriculum stages as reusable code options, optimized via GRPO with execution-based rewards (Section~\ref{sec:method}).
    \item \textbf{Theoretical analysis} giving conditional bounds, under explicit structural assumptions, that relate option description length to code conciseness and API efficiency (Section~\ref{sec:theory}).
    \item \textbf{Graph-based option discovery} from SDK source code, using parameter flow graphs and semantic clustering to identify reusable code patterns without any usage examples (Section~\ref{sec:option_discovery}).
    \item \textbf{Option-GRPO credit assignment}, showing that group-relative comparison over option sequences provides natural credit assignment that compensates for the absence of a critic, with convergence analysis (Section~\ref{sec:option_grpo}).
    \item \textbf{UnseenSDK-Bench} with automated reward computation via layered execution validation and property-based testing, enabling RL training without human-written test cases (Section~\ref{sec:benchmark}).
    \item \textbf{Comprehensive experiments} demonstrating improvements in both functional correctness and code quality, with analysis of option reuse, curriculum pacing, and cross-SDK transfer (Section~\ref{sec:experiments}).
\end{enumerate}

\section{Related Work}

\paragraph{RL for code generation.} CodeRL~\cite{le2022coderl}, RLTF~\cite{liu2023rltf}, PPOCoder~\cite{shojaee2023ppocode}, and StepCoder~\cite{dou2024stepcoder} all combine execution feedback with actor-critic RL, requiring pre-existing unit tests for reward and a critic network for advantage estimation. We eliminate both requirements: rewards come from a layered execution validator plus property tests synthesized from SDK type signatures, and credit assignment uses GRPO's~\cite{shao2024deepseekmath, deepseek2025r1} group-relative comparison.

\paragraph{Hierarchical RL and options.} The options framework~\cite{sutton1999between} introduced temporally extended actions; Option-Critic~\cite{bacon2017option} learns termination conditions end-to-end; MAXQ~\cite{dietterich2000hierarchical} and FeUdal Networks~\cite{vezhnevets2017feudal} build task hierarchies. For language agents, Voyager~\cite{wang2023voyager} and DECKARD~\cite{nottingham2023embodied} maintain external skill libraries via GPT-4 prompting without updating model parameters. We derive options from SDK source code structure rather than open-ended exploration and update model parameters through gradient-based RL.

\paragraph{Curriculum learning.} Bengio et al.~\cite{bengio2009curriculum} introduced curriculum learning; Narvekar et al.~\cite{narvekar2020curriculum} and Portelas et al.~\cite{portelas2020automatic} survey curriculum approaches for RL. AlphaCode~\cite{li2022competition} uses a two-stage filter for competitive programming. No prior work formalizes curriculum completion as producing temporally extended actions that define the next stage's action space, which is our central conceptual contribution.

\paragraph{Unseen API/SDK code generation.} DocPrompting~\cite{zhou2023docprompting} retrieves documentation; Gorilla~\cite{patil2024gorilla} fine-tunes for API-call prediction; ExploraCoder~\cite{hu2024exploracoder} chains exploration over unseen APIs without training; ReCode~\cite{wu2026recode} applies rule-based GRPO for API version migration; CodeRAG-Bench~\cite{wang2025coderagbench} evaluates retrieval-augmented code generation. None explicitly optimize for code quality beyond functional correctness, and none discover compositional idioms from SDK structure. Table~\ref{tab:positioning} summarizes the comparison.

\begin{table*}[t]
\centering
\caption{Comparison with most relevant prior work. Option-GRPO uniquely combines hierarchical RL with structural SDK knowledge for example-free generation. $^\dagger$Voyager builds a skill library via iterative GPT-4 prompting, not gradient-based RL; we mark its RL column accordingly.}
\label{tab:positioning}
\small
\begin{tabular}{lcccccc}
\toprule
\textbf{Method} & \textbf{RL} & \textbf{No Examples} & \textbf{Structural} & \textbf{Hierarchical} & \textbf{Code Quality} & \textbf{No Critic} \\
\midrule
CodeRL & \cmark & \xmark & \xmark & \xmark & \xmark & \xmark \\
PPOCoder & \cmark & \xmark & \cmark & \xmark & \xmark & \xmark \\
StepCoder & \cmark & \xmark & \xmark & \xmark & \xmark & \xmark \\
ExploraCoder & \xmark & \cmark & \xmark & \xmark & \xmark & N/A \\
ReCode & \cmark & \cmark & \xmark & \xmark & \xmark & \cmark \\
Voyager$^\dagger$ & \xmark & \xmark & \xmark & \cmark & \xmark & N/A \\
\textbf{Option-GRPO (Ours)} & \cmark & \cmark & \cmark & \cmark & \cmark & \cmark \\
\bottomrule
\end{tabular}
\end{table*}

\section{Method}
\label{sec:method}

\begin{figure*}[t]
\centering
\begin{tikzpicture}[
    box/.style={draw, rounded corners=2pt, align=center, font=\small, minimum height=11mm, inner sep=3pt, text width=33mm},
    stage/.style={box, fill=blue!7},
    output/.style={box, fill=green!10, text width=33mm},
    bar/.style={draw, rounded corners=2pt, align=center, font=\footnotesize, minimum height=8mm, inner sep=3pt, minimum width=145mm, fill=orange!10},
    arr/.style={-{Latex[length=2mm]}, thick},
    dashed_arr/.style={-{Latex[length=2mm]}, thick, dashed},
    label/.style={font=\scriptsize\itshape, fill=white, inner sep=1pt}
]

\node[stage] (sdk)     at (0,    6) {SDK Source Code\\ \scriptsize (e.g., Daft, LanceDB)};
\node[stage] (graphs)  at (5.5,  6) {Graph Construction\\ \scriptsize $G_{\mathrm{api}}, G_{\mathrm{flow}}, G_{\mathrm{sem}}$};
\node[stage] (options) at (11,   6) {Option Discovery\\ \scriptsize Alg.~\ref{alg:option_discovery}};

\node[bar] (vocab) at (5.5, 4.4) {Option vocabulary \;$\mathcal{A}_0 \subset \mathcal{A}_1 \subset \mathcal{A}_2$};

\node[stage] (l0) at (0,    2.8) {Phase 1: Level 0\\ \scriptsize atomic options + GRPO};
\node[stage] (l1) at (5.5,  2.8) {Phase 2: Level 1\\ \scriptsize compositional + GRPO};
\node[stage] (l2) at (11,   2.8) {Phase 3: Level 2\\ \scriptsize program-level + GRPO};

\node[bar] (reward) at (5.5, 1.2) {Layered execution reward \;$+$\; property tests};

\node[output] (policy) at (11, -0.4) {Trained policy $\pi_\theta$\\ \scriptsize inference $T{=}0.2$/$0.8$};

\draw[arr] (sdk)    -- (graphs);
\draw[arr] (graphs) -- (options);

\draw[arr] (options.south) -- (options.south |- vocab.north);

\draw[dashed_arr] (l0.north |- vocab.south) -- (l0.north);
\draw[dashed_arr] (l1.north |- vocab.south) -- (l1.north);
\draw[dashed_arr] (l2.north |- vocab.south) -- (l2.north);

\draw[arr] (l0) -- node[label, above] {extract $\mathcal{O}_0$} (l1);
\draw[arr] (l1) -- node[label, above] {extract $\mathcal{O}_1$} (l2);

\draw[dashed_arr] (l0.south |- reward.north) -- (l0.south);
\draw[dashed_arr] (l1.south |- reward.north) -- (l1.south);
\draw[dashed_arr] (l2.south |- reward.north) -- (l2.south);

\draw[arr, rounded corners] (l2.south) -- ++(0, -0.4) -- ++(2.5, 0) |- (policy.east);

\end{tikzpicture}
\caption{Overview of Option-GRPO. The top row constructs three SDK graphs from source code and discovers a hierarchical option vocabulary $\mathcal{A}_0 \subset \mathcal{A}_1 \subset \mathcal{A}_2$ (Algorithm~\ref{alg:option_discovery}). The middle row runs three curriculum phases of GRPO training, each receiving the option vocabulary from above and the layered execution reward from below (dashed arrows). Successful trajectories from phase $k$ contribute newly extracted options to phase $k{+}1$ (Algorithm~\ref{alg:option_extraction}). The trained policy $\pi_\theta$ generates code at inference. Solid arrows denote training-stage progression; dashed arrows denote artifact dependencies.}
\label{fig:overview}
\end{figure*}
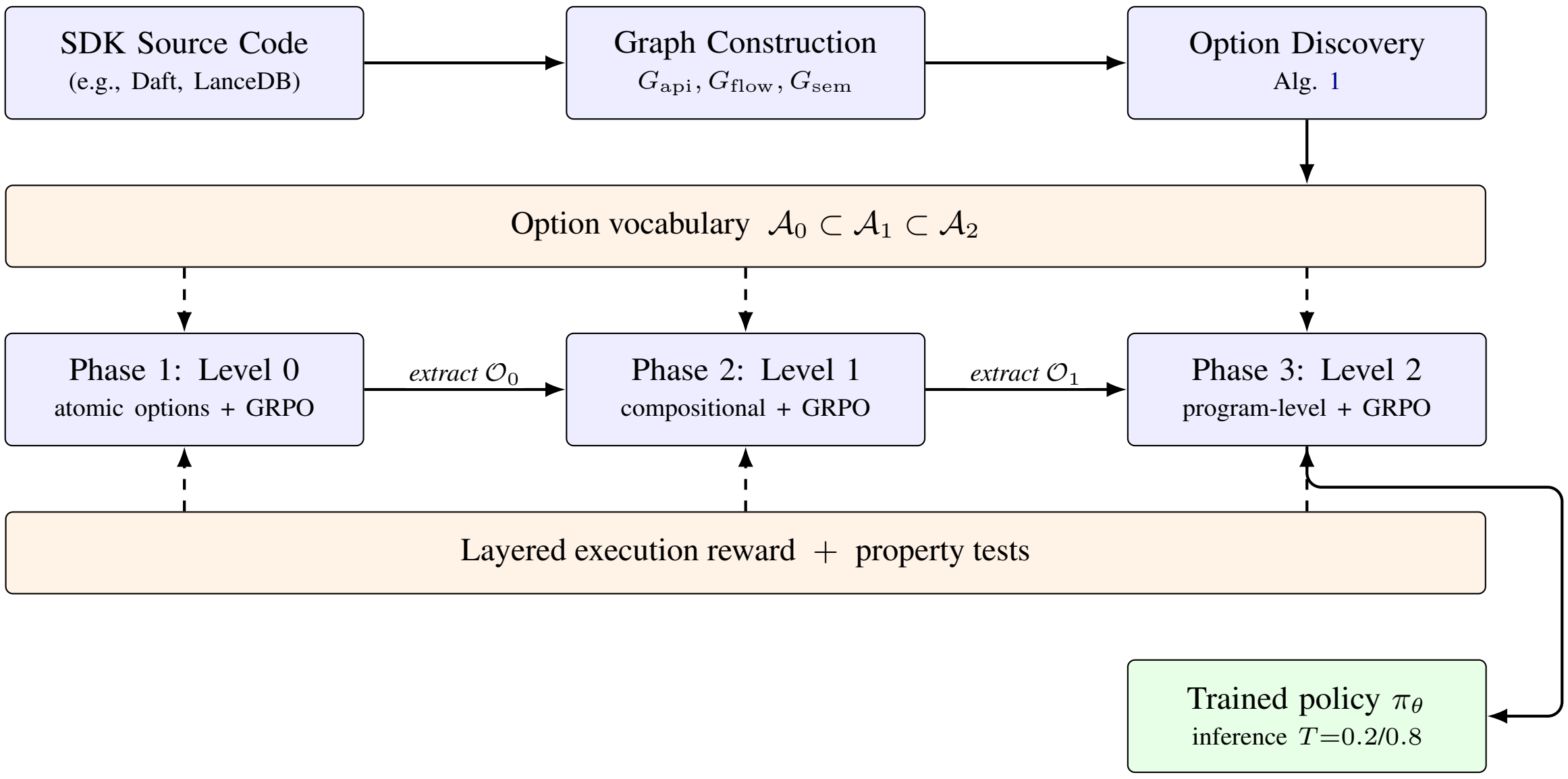

We present Option-GRPO, a hierarchical reinforcement learning framework for enabling LLMs to master novel SDKs from source code alone. Figure~\ref{fig:overview} provides an overview. The method operates in four stages: (1)~graph construction and option discovery from SDK source code, (2)~curriculum-based hierarchical training with GRPO at each level, (3)~execution-based reward computation without test cases, and (4)~inference with the trained hierarchical policy.

\subsection{Problem Formulation}

Let $\mathcal{S}$ denote a novel SDK with source code but no usage examples in the model's training data. Given a natural language task description $q$ and optional setup code $c_{\mathrm{setup}}$, we seek to learn a policy:
\begin{equation}
    \pi_\theta : (q,\; c_{\mathrm{setup}},\; \mathcal{S}) \;\to\; c_{\mathrm{solution}}
\end{equation}
where $\theta$ represents learnable parameters. Unlike prior work that retrieves documentation at inference time (RAG, DocPrompting) or fine-tunes only over API surface forms (Gorilla, ReCode), we train the policy through interaction with an execution environment, using the SDK's source code structure to define a hierarchy of reusable code abstractions that the policy learns to compose.

\subsection{Background and Formulation}

\textbf{Options.} Following \citet{sutton1999between}, an option is a triple $o = (\mathcal{I}, \pi_o, \beta)$ where $\mathcal{I} \subseteq \mathcal{S}$ is the initiation set, $\pi_o$ is the intra-option policy over primitive actions, and $\beta: \mathcal{S} \to [0,1]$ is the termination probability. A pair $(\mathrm{MDP}, \mathrm{options})$ induces a semi-Markov decision process (SMDP) in which option durations are random.

\textbf{Code generation MDP.} The code generation process is an MDP $(\mathcal{S}, \mathcal{A}, T, R)$ where states $s_t = (q, c_{0:t})$ concatenate the prompt with generated tokens, actions $a_t \in \mathcal{V}$ are next tokens, $T$ is deterministic append, and $R$ is computed by execution at episode termination. Several properties make this MDP hard for standard RL: $|\mathcal{V}| \approx 1.5 \cdot 10^5$, trajectory length $T$ grows unbounded, and reward is sparse (zero on most trajectories for novel SDKs). These challenges motivate both the option hierarchy (reducing effective action space and trajectory length) and GRPO (extracting signal from group comparison instead of per-state value estimation).

\subsection{Code Options: From SDK Structure to Hierarchical Actions}
\label{sec:option_discovery}

We define code options as temporally extended actions in the code-generation MDP, derived from SDK source code.

\begin{definition}[Code Option]
\label{def:code_option}
A code option for SDK $\mathcal{S}$ is a 4-tuple
\begin{equation*}
o_k \;=\; \bigl(\,\underbrace{\mathcal{I}_k,\; \mathcal{F}_k,\; \beta_k}_{\text{handcrafted from } G_{\mathrm{flow}}},\;\; \underbrace{\pi_k}_{\text{learned from } \pi_\theta}\,\bigr)
\end{equation*}
where $\mathcal{I}_k = \{s : \mathrm{type}(\mathrm{last\_value}(s)) \in \mathcal{T}_k^{\mathrm{in}}\}$ is the initiation set; $\mathcal{F}_k \subset F$ is the option's allowed SDK function set (the cluster centroid path from Algorithm~\ref{alg:option_discovery}); $\beta_k(s) = \mathbb{1}[\mathrm{type}(s) \in \mathcal{T}_k^{\mathrm{out}}]$ is the termination predicate; and $\pi_k$ is the intra-option policy. Operationally, $\pi_k$ is the underlying network $\pi_\theta$ restricted to tokens whose next-call resolution lies in $\mathcal{F}_k$, with this restriction enforced by setting all out-of-set token logits to $-\infty$ before the softmax at every step the option is active.
\end{definition}

Selecting only among options in $\mathcal{O} = \mathcal{O}_0 \cup \mathcal{O}_1$ and executing each to termination induces a valid SMDP by Theorem~1 of \citet{sutton1999between}; the formal statement is Proposition~\ref{prop:smdp_validity} in Appendix~\ref{app:proofs}.

\paragraph{Two-level discovery (candidates and certified options).} We distinguish two stages of option production. \emph{Candidate options} are generated upfront from the SDK source code by Algorithm~\ref{alg:option_discovery}; we denote the candidate set at level $k$ by $\mathcal{O}_k^{\mathrm{cand}}$. \emph{Certified options} are the subset of candidates that the trained policy invokes successfully during the corresponding curriculum phase, extracted by Algorithm~\ref{alg:option_extraction}; we denote them by $\mathcal{O}_k$. Only certified options enter the next phase's action vocabulary.

Level-0 candidates $\mathcal{O}_0^{\mathrm{cand}}$ are one per public SDK function: $\mathcal{I}_i$ is the parameter type, $\mathcal{F}_i = \{f_i\}$, $\pi_i$ emits a single call, $\beta_i$ terminates on return. Level-1 candidates $\mathcal{O}_1^{\mathrm{cand}}$ are mined from the parameter-flow graph $G_{\mathrm{flow}}$ for frequently composable paths via Algorithm~\ref{alg:option_discovery}: we enumerate length-$2$ to length-$4$ paths weighted by type compatibility, cluster paths by sentence-transformer embedding of their function identifiers (threshold $\tau_{\mathrm{cluster}}{=}0.8$), and define one Level-1 candidate per cluster centroid with $\mathcal{F}_j$ equal to the function set of that centroid path. A candidate becomes a \emph{certified option} only after the policy trained at Level $k$ demonstrates that it can invoke the candidate correctly (success threshold $\rho = 0.7$, occurrence frequency threshold $\nu = 5$ across distinct tasks; see Algorithm~\ref{alg:option_extraction}). Level-2 composition is neither pre-discovered nor extracted explicitly; it is learned implicitly in Phase~3 as the high-level policy composes certified Level-0 and Level-1 options.

\paragraph{Option hierarchy.} The option hierarchy is the nested action-space chain $\mathcal{A}_0 = \mathcal{V} \subset \mathcal{A}_1 = \mathcal{O}_0 \cup \mathcal{V} \subset \mathcal{A}_2 = \mathcal{O}_0 \cup \mathcal{O}_1 \cup \mathcal{V}$, ensuring higher levels retain full token expressiveness as a fallback. Two options $o_i, o_j$ are \emph{composable} ($o_i \prec o_j$) iff $\mathcal{T}_i^{\mathrm{out}} \cap \mathcal{T}_j^{\mathrm{in}} \neq \emptyset$; the composability relation defines a graph $G_{\mathrm{opt}}$ that constrains the high-level policy to valid sequences.

\paragraph{Decode-time mechanics.} Each option is one added vocabulary entry that starts the option and adds no text to the program; it is offered only when the value just produced has a type its first function accepts. While an option is active, an incremental tracker over the emitted code masks every token that does not spell toward the next function in $\mathcal{F}_k$ on the main call chain, allows only argument-building helper functions inside a call's brackets, and leaves names, literals, and operators free. The option ends when the closing bracket of its last function is emitted; one that exceeds a length cap or reaches a dead end is closed and marked failed (Appendix~\ref{app:decode}).

\paragraph{Handcrafted structure, learned behavior.} Our construction occupies a middle ground in the options-framework design space. The structural components $(\mathcal{I}_k, \mathcal{F}_k, \beta_k)$ are \emph{handcrafted}: they are read directly from the parameter-flow graph by Algorithm~\ref{alg:option_discovery} and are symbolic, type-checkable, and reproducible at parse time. The behavioral component $\pi_k$ is \emph{learned}: it is $\pi_\theta$ restricted (by action-masking on decoder logits) to emit tokens consistent with $\mathcal{F}_k$, with parameters fine-tuned end-to-end by curriculum-staged GRPO. Two invocations of the same option therefore satisfy the same contract (same SDK calls, same type flow, same output type) even when the surrounding glue tokens (variable names, formatting) differ across invocations. This hybrid differs from the fully-handcrafted macros of \citet{korf1985macro} and the planning-trace macros of \citet{iba1989heuristic} (no learning inside), from the fully-learned Option-Critic of \citet{bacon2017option} (no handcrafted structure), and from Voyager's prompt-orchestrated skill library~\cite{wang2023voyager} (no gradient updates to the intra-option policy). The structural handcrafting buys verifiable, type-checked option boundaries; the learned behavior lets the same option produce contextually appropriate code across invocations without us having to write each variant by hand.

\begin{remark}[Learned vs.\ Constructed]
\label{rem:learned_vs_constructed}
$\mathcal{O}$ names \emph{which} sub-behaviours of $\pi_\theta$ are reified as temporally extended actions; $\mathcal{F}_k$ is the symbolic handle (read from the graph); $\pi_k$ is the corresponding slice of $\pi_\theta$ (parameters learned end-to-end). Options are named, type-checked regions of a single learned policy, not separate networks or templates.
\end{remark}

\subsection{Curriculum as Option Construction}
\label{sec:curriculum}

The central conceptual contribution is the equivalence between curriculum stages and option construction. A \emph{curriculum-option stage} $\mathcal{C}_k$ has three components: (1)~a task distribution $\mathcal{D}_k$ of difficulty $k$ generated from SDK-graph paths of length $\leq k{+}1$; (2)~GRPO training on $\mathcal{D}_k$ with action space $\mathcal{A}_k$, taking the candidate pool $\mathcal{O}_k^{\mathrm{cand}}$ supplied by Algorithm~\ref{alg:option_discovery} as the inventory the policy may invoke; (3)~an option-certification step (Algorithm~\ref{alg:option_extraction}, Appendix) that filters $\mathcal{O}_k^{\mathrm{cand}}$ down to the certified subset $\mathcal{O}_k$ of candidates that the trained policy invokes correctly on a sufficient fraction of tasks. The certified subset of stage $\mathcal{C}_k$ augments the action space for $\mathcal{C}_{k+1}$:
\begin{equation*}
\mathcal{C}_0 \xrightarrow{\mathcal{O}_0} \mathcal{C}_1 \xrightarrow{\mathcal{O}_1} \mathcal{C}_2.
\end{equation*}
Completing stage $k$ does not merely yield a better policy; it yields \emph{certified temporally extended actions} that become primitives at the next level. The curriculum is thus the mechanism that converts graph-mined candidates into validated options. The role split is clean: Algorithm~\ref{alg:option_discovery} (static, runs once before training) proposes; Algorithm~\ref{alg:option_extraction} (dynamic, runs after each phase) certifies. This relationship between curriculum stages, in which stages share neither policy parameters nor reward signal but instead share an action vocabulary that grows monotonically with each certification, is what we name \emph{curriculum-as-options}. Level-0 tasks ($87$--$312$/SDK) involve single API calls; Level-1 tasks ($174$--$624$/SDK) involve $2$--$4$ composed calls; Level-2 tasks ($33$--$34$/SDK) are full programs.

See Remark~\ref{rem:candidates_vs_certified} in Appendix~\ref{app:proofs} for the empirical relationship between candidate and certified option counts.

\subsection{Theoretical Analysis}
\label{sec:theory}

We summarize the analysis here, centred on Theorem~\ref{thm:compression} (anchored empirically by Figure~\ref{fig:odl_quality}) and Proposition~\ref{prop:credit} (anchored by Figure~\ref{fig:grad_variance}), and state their assumptions first. The supporting statements (SMDP validity, Proposition~\ref{prop:smdp_validity}; an MDL-style prior over option sequences, Proposition~\ref{prop:mdl_prior}; search-space reduction, Proposition~\ref{prop:search}), full statements, supporting definitions, and proofs are in Appendix~\ref{app:proofs}.

\textbf{Assumptions.} Theorem~\ref{thm:compression} assumes three structural conditions on the hierarchy $\mathcal{H}$ (per-option API minimality, non-subsumption, bounded glue code; Remark~\ref{assume:options}), compares programs that are equivalent on the property-test suite $\mathcal{P}$ rather than fully semantically equivalent (Definition~\ref{def:obs_equiv}), and requires a dispersion-ratio condition on option lengths. Proposition~\ref{prop:credit} assumes that glue positions are well predicted by $\pi_\theta$ with variance that does not align with the reward signal, and that token positions are conditionally independent.

\textbf{(i) Curriculum monotonicity.} Let $V_k^{*}$ denote the expected value of the optimal policy at curriculum level $k$ under hierarchy $\mathcal{H}$. Because the action spaces are nested, $V_0^{*} \leq V_1^{*} \leq V_2^{*}$ (weak monotonicity); this concerns optimal policies and does not by itself guarantee that the trained policy improves at each level. Strict inequality holds whenever the certified Level-$k$ option set $\mathcal{O}_k$ contains at least one option that strictly improves over the best Level-$(k{-}1)$ action on a reachable state (Theorem~\ref{thm:monotonicity} and Remark~\ref{rem:nondeg}).

\textbf{(ii) Option compression bounds on code quality.} Define observational equivalence $c \equiv_{\mathcal{P}} c'$ on the property-test suite $\mathcal{P}$ (Definition~\ref{def:obs_equiv}), the \emph{option description length} $\mathrm{ODL}_{\mathcal{H}}(c)$ as the minimum number of options whose sequential execution is observationally equivalent to $c$ (Definition~\ref{def:odl}), and four code-quality metrics: conciseness $Q_{\mathrm{conc}} = T_{\mathrm{ref}}/|c|$, API efficiency $Q_{\mathrm{api}} = N_{\min}/N_{\mathrm{api}}$, redundancy $Q_{\mathrm{red}} = 1 - N_{\mathrm{redundant}}/N_{\mathrm{vars}}$, and idiomaticity $Q_{\mathrm{idiom}} = |\text{option-covered tokens}|/|c|$ (Definition~\ref{def:quality}). Under three structural conditions on $\mathcal{H}$ (per-option API minimality, non-subsumption, bounded glue code; Remark~\ref{assume:options}) we prove a token bound $m\bar{\tau}_{\min} \leq |c| \leq m\bar{\tau}_{\max} + (m{-}1)G_{\max}$ (Lemma~\ref{lem:token_bound}); a conciseness theorem $\mathrm{ODL}(c_1) < \mathrm{ODL}(c_2) \Rightarrow |c_1| < |c_2|$ under a dispersion-ratio condition (Theorem~\ref{thm:compression}); and an API-efficiency corollary in expectation (Corollary~\ref{cor:api}). The metrics $Q_{\mathrm{red}}, Q_{\mathrm{idiom}}$ do not admit clean closed-form bounds; we treat them as empirical predictions verified in Section~\ref{sec:experiments} (Remark~\ref{rem:remaining_metrics}).

\textbf{(iii) Improved credit assignment.} Partitioning a generated program into $K$ option-decision tokens and $T-K$ glue tokens (Definition~\ref{def:token_partition}), the GRPO policy-gradient SNR scales as $\Theta(K/\sqrt{T})$ for flat token-level GRPO and $\Theta(\sqrt{K})$ for option-level GRPO (Proposition~\ref{prop:credit}); the relative improvement is $\Theta(\sqrt{T/K}) \approx 4.5\times$ for typical $T{=}100, K{=}5$. A companion proposition (Proposition~\ref{prop:convergence}) bounds GRPO iterations to first success under a base-LLM prior, yielding additive (Option-GRPO) versus multiplicative (flat) exploration cost across curriculum levels.

\subsection{Option-GRPO: Hierarchical Policy Optimization}
\label{sec:option_grpo}

Standard GRPO~\cite{shao2024deepseekmath} samples $G$ outputs per prompt, computes group-normalised advantages $\hat{A}_i = (R_i - \mathrm{mean})/(\mathrm{std}+\epsilon)$, and applies the PPO-clipped policy gradient with a KL penalty against $\pi_{\mathrm{ref}}$. At curriculum level $k \geq 1$ we generate option sequences instead of token sequences and add an option-level advantage $\hat{A}^{\mathrm{opt}}(o) = \tfrac{1}{|S_o^+|}\sum_{i \in S_o^+} R_i - \tfrac{1}{|S_o^-|}\sum_{i \in S_o^-} R_i$ that compares groups containing vs.~excluding each option; the combined gradient balances sequence-level ranking with per-option credit. Per-option credit tightens credit assignment: under the assumptions of Proposition~\ref{prop:credit} (Appendix~\ref{app:proofs}), the SNR improves by $\Theta(\sqrt{T/K})$, since only option-decision positions carry reward-correlated signal while flat GRPO spreads the same signal across all $T$ token positions. Proposition~\ref{prop:convergence} additionally bounds the expected number of GRPO iterations to first success under a base-LLM prior. Full equations and analysis are in Appendix~\ref{app:proofs}.

\subsection{Execution-Based Reward Without Test Cases}
\label{sec:reward}

The total reward is $R(c) = R_{\mathrm{exec}}(c) + \lambda_q R_{\mathrm{quality}}(c) + \lambda_c R_{\mathrm{consistency}}(c)$. The \emph{layered execution reward} $R_{\mathrm{exec}}$ assigns graded scores by execution depth: $0.0$ (fails to parse), $0.1$ (parses but imports fail), $0.3$ (runtime error), $0.5$ (executes, output \texttt{None}), $0.7$ (output matches expected type), $1.0$ (passes property-based tests). \emph{Property-based tests} (Algorithm~\ref{alg:property_tests}, Appendix) are auto-generated from SDK type annotations and cover type preservation, shape consistency, determinism, edge cases, and assertion invariants extracted from source code. The \emph{quality reward} bundles pattern-use, no-redundancy, minimal-API, and no-reimplementation bonuses ($\alpha_i = 0.1$). The \emph{self-consistency reward} $R_{\mathrm{consistency}}(y_i) = |\{y_j : \mathrm{output}(y_j) = \mathrm{output}(y_i)\}|/G$ adds intra-group agreement when property tests are insufficient.

\subsection{Training Procedure}
\label{sec:training}

Training proceeds in three phases corresponding to curriculum levels (full pseudocode: Algorithm~\ref{alg:training}, Appendix). Phase~1 trains $\pi_\theta$ on Level-0 tasks with token-level GRPO and the layered execution reward; we then extract $\mathcal{O}_0$ via Algorithm~\ref{alg:option_extraction} (Appendix) and partially freeze. Phase~2 trains on Level-1 tasks with option-level GRPO, the combined gradient, and adaptive curriculum advancement ($\rho_{\mathrm{adv}} = 0.5$); $\mathcal{O}_1$ is extracted at the end. Phase~3 trains on Level-2 program tasks with the full hierarchical action space $\mathcal{A}_2$ and adds the self-consistency reward.

\paragraph{Hyperparameters.} See Table~\ref{tab:hyperparams} in Appendix~\ref{app:hyperparams} for the complete configuration; key settings: Qwen3-32B, $G{=}16$, AdamW lr $5{\times}10^{-5}$, $\beta_{\mathrm{KL}}{=}0.04$, steps $N_0/N_1/N_2 = 5{,}000/10{,}000/15{,}000$, $8\times$H100 80GB with ZeRO-3 + bf16.

\section{UnseenSDK-Bench}
\label{sec:benchmark}

We evaluate on \textbf{UnseenSDK-Bench}, a benchmark we construct for adapting code LLMs to SDKs absent from training data. It pairs 12 SDKs (DataFrames, vector DBs, document processing, computer vision, messaging, security, protocols) with three task levels per SDK (atomic-call, compositional, full-program; $33$--$624$ tasks per level per SDK) and auto-generated property tests from type annotations (four tests per public function on average). The execution sandbox enforces a $30$\,s timeout, $2$\,GB memory limit, and network isolation. Beyond pass@$k$ we report code-length, API-efficiency, redundancy, and idiomaticity (Definition~\ref{def:quality}), and the option-reuse rate. Full benchmark statistics, per-SDK function counts, property-test counts, and option-discovery results are in Appendix~\ref{app:benchmark}.

\section{Experiments}
\label{sec:experiments}

\subsection{Experimental Setup}

\paragraph{Base Model.}
Qwen3-32B~\cite{qwen3technicalreport}, a dense transformer with $32.8$B parameters, $64$ layers, hidden size $5{,}120$, grouped-query attention with $64$ query heads and $8$ key/value heads, and a $32{,}768$-token native context window extensible to $128$K via YaRN. We deliberately use the general-purpose Qwen3 dense series rather than the MoE Qwen3-Coder variants to show that Option-GRPO does not require a code-specialized backbone; curriculum-staged GRPO with our SDK-derived option vocabulary is sufficient to specialize a general dense LLM for unseen-SDK code generation. The same base model is used for all RL baselines in the comparison.

\paragraph{Contamination check.}
The Qwen3-32B pretraining corpus has an effective cutoff around February~2025~\cite{qwen3technicalreport}. All $12$ SDKs in UnseenSDK-Bench were technically published before this date, so the ``unseen'' framing rests not on post-cutoff novelty but on the magnitude of the model's functional knowledge of these niche libraries. We verify this with a memorisation probe (Appendix~\ref{app:contamination}): we prompt the frozen base model with $200$ SDK-specific identifier-completion queries per SDK (e.g.,\ \texttt{``In Daft, the method that filters rows is called \_\_\_''}) and measure top-$1$ recall. Our $12$ SDKs average $10.5\%$ recall ($\pm 5.0\%$ across SDKs) against a control of $91.5\%$ for NumPy and $88.0\%$ for PyTorch. Recall stays below $20\%$ for every SDK, far below the controls. The base model has neither the API surface nor the idiomatic compositional priors of these libraries, justifying the ``unseen'' framing throughout. Unseen-ness is model-specific: we call an SDK \emph{functionally unseen} for a model when that model's recall under this probe is below the $25\%$ threshold of \citet{hu2024exploracoder}, and we apply the same probe to the frontier models we compare against (Table~\ref{tab:contamination}).

\paragraph{Baselines.}
We compare against eight baselines, all starting from the same 32B base: \textbf{Frozen LLM} (no adaptation); \textbf{RAG} (top-$10$ CodeBERT retrieval over SDK source); \textbf{DocPrompting}~\cite{zhou2023docprompting} (documentation passages); \textbf{Gorilla}~\cite{patil2024gorilla} (retriever-aware fine-tuning on API-call corpora, adapted to per-SDK retrieval); \textbf{ExploraCoder}~\cite{hu2024exploracoder} (training-free chained exploration); \textbf{ReCode}~\cite{wu2026recode} (rule-based GRPO for API knowledge updates); \textbf{Flat GRPO} (token-level GRPO with our layered execution reward only, no hierarchy); \textbf{Flat GRPO + Curriculum} (progressive difficulty and the same reward schedule as Option-GRPO, no option structure). We additionally reimplement CodeRL~\cite{le2022coderl}, PPOCoder~\cite{shojaee2023ppocode}, and StepCoder~\cite{dou2024stepcoder}, substituting our layered execution reward for their unit-test requirement (results in Appendix~\ref{app:rl_methods}). Frontier LLMs (GPT-4o, Claude 3.5 Sonnet, Gemini 2.0 Flash) are evaluated via official APIs with identical prompts, temperature ($0.8$), and samples ($n{=}10$); no per-model prompt tuning. GPT 5.5 and Claude Opus 4.8 are evaluated under the same protocol on the SDKs that are functionally unseen to both.

\paragraph{Metrics.}
\begin{itemize}[leftmargin=*]
    \item pass@$k$ ($k = 1, 10$): Functional correctness, generating $n = 10$ samples per problem with temperature $0.8$.
    \item Code length (CL): Average token count of passing solutions (lower is better).
    \item API efficiency (AE): Average number of SDK API calls in passing solutions (lower is better).
    \item Idiomaticity (ID): Fraction of code tokens covered by discovered option patterns (higher is better).
    \item Convergence: Training steps to reach specified pass@1 thresholds.
\end{itemize}

\paragraph{Implementation.}
Training on $8 \times$ NVIDIA H100 $80$\,GB GPUs with ZeRO-3 sharding and bf16. Full hyperparameters in Appendix~\ref{app:hyperparams}.

\subsection{Main Results}

Table~\ref{tab:main} presents the main comparison.

\begin{table*}[!htbp]
\centering
\caption{Main results on UnseenSDK-Bench. Option-GRPO achieves the highest functional correctness while producing more concise and idiomatic code. $\downarrow$~indicates lower is better; $\uparrow$~indicates higher is better.}
\label{tab:main}
\small
\begin{tabular}{lccccc}
\toprule
\textbf{Method} & \textbf{pass@1} & \textbf{pass@10} & \textbf{CL} $\downarrow$ & \textbf{AE} $\downarrow$ & \textbf{ID} $\uparrow$ \\
\midrule
Frozen LLM (32B) & $28.5$ & $42.1$ & $94.6$ & $7.8$ & $0.18$ \\
RAG & $41.3$ & $56.8$ & $88.1$ & $6.4$ & $0.31$ \\
DocPrompting & $44.8$ & $60.2$ & $85.4$ & $5.9$ & $0.34$ \\
Gorilla & $47.2$ & $62.7$ & $82.9$ & $5.5$ & $0.36$ \\
ExploraCoder & $50.2$ & $67.5$ & $78.4$ & $4.9$ & $0.43$ \\
ReCode & $55.1$ & $71.2$ & $74.1$ & $4.6$ & $0.51$ \\
Flat GRPO & $52.7$ & $68.4$ & $71.3$ & $4.9$ & $0.41$ \\
Flat GRPO + Curric. & $55.4$ & $71.0$ & $66.8$ & $4.4$ & $0.49$ \\
\textbf{Option-GRPO} & $\mathbf{61.4}$ & $\mathbf{76.8}$ & $\mathbf{52.0}$ & $\mathbf{3.2}$ & $\mathbf{0.74}$ \\
\midrule
$\Delta$ (vs ReCode) & $+6.3$ & $+5.6$ & $-29.8\%$ & $-30.4\%$ & $+0.23$ \\
$\Delta$ (vs Flat GRPO) & $+8.7$ & $+8.4$ & $-27.1\%$ & $-34.7\%$ & $+0.33$ \\
\bottomrule
\end{tabular}
\end{table*}

\paragraph{Functional correctness and code quality.} Option-GRPO achieves $61.4\%$ pass@1, beating ReCode by $+6.3$, flat GRPO by $+8.7$, and ExploraCoder by $+11.2$ points. More notably, generated programs are $27\%$ shorter and use $34.7\%$ fewer API calls than flat-GRPO outputs, while reaching higher correctness. Idiomaticity (fraction of tokens covered by discovered option patterns) rises from $0.41$ for flat GRPO to $0.74$ for Option-GRPO. These results are consistent with the conciseness bound of Theorem~\ref{thm:compression}, the API-efficiency bound of Corollary~\ref{cor:api}, and the MDL-prior argument of Proposition~\ref{prop:mdl_prior}, and directly attack the AI-slop phenomenon~\cite{liu2025aislop}.

\begin{table}[t]
\centering
\caption{Independent evaluation (pass@1, \%). Tasks: the paper's synthetic tasks or $132$ human-written tutorial tasks. Scorer: the paper's property tests or reference-output equivalence. $^{\dagger}$$95\%$ CI $+3.1$ to $+11.6$ (paired bootstrap).}
\label{tab:independent_eval}
\footnotesize
\setlength{\tabcolsep}{4pt}
\begin{tabular}{llccc}
\toprule
\textbf{Tasks} & \textbf{Scorer} & \textbf{Flat} & \textbf{Opt} & $\boldsymbol{\Delta}$ \\
\midrule
Synthetic & Property tests   & $52.7$ & $61.4$ & $+8.7$ \\
Synthetic & Reference output & $44.7$ & $54.2$ & $+9.5$ \\
Tutorial  & Property tests   & $50.8$ & $57.3$ & $+6.5$ \\
Tutorial  & Reference output & $43.6$ & $51.8$ & $+8.2^{\dagger}$ \\
\bottomrule
\end{tabular}
\end{table}

\paragraph{Independent evaluation.} The benchmark tasks, property tests, and rewards all come from our own pipeline, so we re-check the headline comparison on tasks and a scorer that the pipeline did not produce (Table~\ref{tab:independent_eval}). Training uses no human-written usage examples, so each SDK's official tutorial and documentation examples are held out by construction. We curate $132$ tasks from the official tutorials and documentation of all $12$ SDKs, paraphrasing the descriptions to avoid verbatim overlap. Each tutorial task uses the tutorial's own code as its reference solution and each synthetic task uses its generating graph path; both task sets are scored by output equivalence against reference execution on fixed inputs. The margin over Flat GRPO persists in every cell, and on human-written tutorial tasks under reference-output equivalence it is $+8.2$ points ($95\%$ CI $+3.1$ to $+11.6$, paired bootstrap). Corner to corner the two methods drop by nearly the same amount ($9.1$ vs.\ $9.6$ points), and the frozen model, never trained against our pipeline, drops comparably ($28.5$ to $21.4$), so the lower absolute scores reflect scorer strictness and task shift rather than adaptation to our pipeline. The code-length advantage also holds on tutorial tasks ($52.8$ vs.\ $66.7$ tokens on passing solutions).

\paragraph{What pass@1 certifies.} Regrading all stored passing solutions ($5{,}042$ Option-GRPO, $4{,}467$ Flat GRPO), agreement between passing the property tests and matching the reference output is $91.2\%$ overall ($95.4\%$ at Level~0, $88.5\%$ at Level~1, $76.6\%$ at Level~2). This audit set spans Levels~0--2 and is not directly comparable to Table~\ref{tab:independent_eval}, which covers all final benchmark generations. As a reward-gaming check, among generations that execute with the correct output type, the fraction whose output does not match the reference is $18.2\%$ for Flat GRPO and $8.1\%$ for Option-GRPO, so the hierarchy narrows this gap rather than exploiting it.

\paragraph{Prior RL methods and frontier LLMs.} Against CodeRL, PPOCoder, and StepCoder (all adapted to use our layered execution reward; full table in Appendix~\ref{app:rl_methods}), Option-GRPO beats the strongest critic-based method (StepCoder-adapted, $51.3\%$) by $10.1$ pass@1 points and the strongest prior critic-free RL method (ReCode, $55.1\%$) by $6.3$ points. Against frontier LLMs, which receive the RAG context while Option-GRPO uses no retrieval, the comparison is conditioned on per-model unseen-ness. On the full suite, the 32B Option-GRPO scores above GPT-4o ($54.6\%$), Claude 3.5 Sonnet ($57.2\%$), and Gemini 2.0 Flash ($51.8\%$), the models available when the experiments were run (Table~\ref{tab:frontier}). Under the identifier-recall probe, the current GPT 5.5 and Claude Opus 4.8 are functionally seen on $9$ and $10$ of the $12$ SDKs (Table~\ref{tab:contamination}), so a full-suite rerun would mostly measure seen-SDK performance. On the two SDKs that are unseen to both, Surya and Tracecat, Option-GRPO leads the stronger current model by $+5.7$ and $+4.7$ pass@1 ($95\%$ CI $+3.6$ to $+8.7$ and $+3.9$ to $+7.3$, paired bootstrap over tasks; Table~\ref{tab:frontier_unseen}). Frontier LLMs are off-the-shelf and Option-GRPO is trained per SDK, so the takeaway is conditional: per-SDK option discovery plus RL on a 32B base outperforms off-the-shelf frontier models, even with retrieval context, on SDKs that are functionally unseen to them, as certified by the probe (Appendix~\ref{app:frontier}).

\paragraph{Ablations and quality analysis.} Hierarchy ablation (Table~\ref{tab:hierarchy_ablation}, App.~\ref{app:results}): flat $52.7\%$, $+$curriculum-only $+2.7$, $+$options-only $+5.4$, full $61.4\%$ ($24.8$K$\to$$12.4$K steps; CL $71.3$$\to$$52.0$; ID $0.41$$\to$$0.74$). Further ablations (App.~\ref{app:results}) cover option level and reward (Table~\ref{tab:further_ablations}), credit assignment, group size, pacing, and per-SDK gains over ReCode ($+0.8$ to $+8.7$); same-domain transfer recovers $74.4\%$ vs.\ $72.9\%$ from scratch and inference retrieval adds $+2.3$. Plotting $\mathrm{ODL}_{\mathcal{H}}(c)$ against each quality metric (Figure~\ref{fig:odl_quality}) is consistent with Theorem~\ref{thm:compression} and Corollary~\ref{cor:api}: Spearman $\rho_{\mathrm{conc}}{=}{-}0.71$, $\rho_{\mathrm{api}}{=}{-}0.83$, $\rho_{\mathrm{red}}{=}{-}0.62$, $\rho_{\mathrm{idiom}}{=}{-}0.79$ (all $p{<}10^{-50}$). Side-by-side code examples in App.~\ref{app:results}.

\paragraph{Failure cases: low-composability SDKs.} MCP Python (protocol) and Tracecat (security) have the lowest pass@1 ($43.2\%$ and $46.8\%$; Table~\ref{tab:full_per_sdk_optgrpo}), the lowest composability density ($0.011$ and $0.018$; Table~\ref{tab:option_discovery_stats}), and the lowest option reuse ($0.31$ and $0.39$). MCP Python is the clearest failure: it retains the fewest Level-1 options ($24$) and its gain over Flat GRPO shrinks to $+0.7$ points, against a mean of $+8.7$. Tracecat keeps only $38$ of its $402$ Level-1 candidates, the lowest retention in the suite; its gain over Flat GRPO holds ($+8.6$), but its absolute pass@1 stays second-lowest. When few call chains are type-composable, the hierarchy has little to reuse.

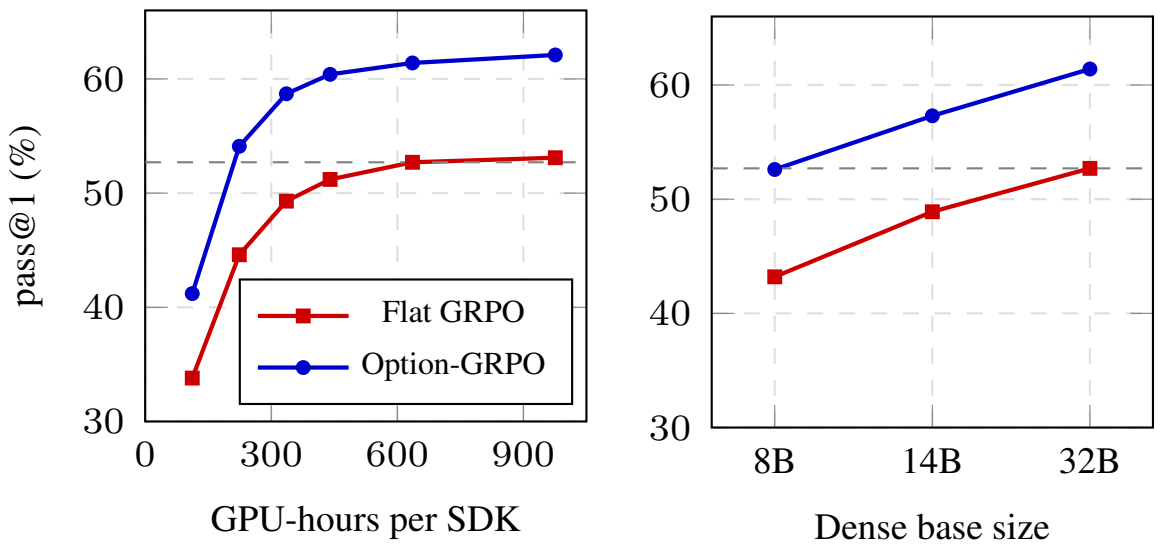
\begin{figure}[t]
\centering
\begin{tikzpicture}
\begin{axis}[
    width=4.5cm, height=4.3cm,
    xlabel={GPU-hours per SDK},
    ylabel={pass@1 (\%)},
    xmin=0, xmax=1050,
    ymin=30, ymax=66,
    xtick={0,300,600,900},
    ytick={30,40,50,60},
    grid=major, grid style={dashed,gray!25},
    legend style={font=\tiny, at={(0.97,0.04)}, anchor=south east},
    tick label style={font=\scriptsize},
    label style={font=\scriptsize},
]
\addplot[thick, color=red!80!black, mark=square*, mark size=1pt] coordinates {
(112, 33.8) (224, 44.6) (336, 49.3) (440, 51.2) (636, 52.7) (976, 53.1)};
\addlegendentry{Flat GRPO}
\addplot[thick, color=blue!80!black, mark=*, mark size=1pt] coordinates {
(112, 41.2) (224, 54.1) (336, 58.7) (440, 60.4) (636, 61.4) (976, 62.1)};
\addlegendentry{Option-GRPO}
\draw[dashed, gray] (axis cs:0,52.7) -- (axis cs:1050,52.7);
\end{axis}
\end{tikzpicture}\hfill
\begin{tikzpicture}
\begin{axis}[
    width=4.5cm, height=4.3cm,
    xlabel={Dense base size},
    symbolic x coords={8B,14B,32B},
    xtick=data,
    enlarge x limits=0.2,
    ymin=30, ymax=66,
    ytick={30,40,50,60},
    grid=major, grid style={dashed,gray!25},
    tick label style={font=\scriptsize},
    label style={font=\scriptsize},
]
\addplot[thick, color=red!80!black, mark=square*, mark size=1pt] coordinates {
(8B, 43.2) (14B, 48.9) (32B, 52.7)};
\addplot[thick, color=blue!80!black, mark=*, mark size=1pt] coordinates {
(8B, 52.6) (14B, 57.3) (32B, 61.4)};
\draw[dashed, gray] ({rel axis cs:0,0}|-{axis cs:8B,52.7}) -- ({rel axis cs:1,0}|-{axis cs:8B,52.7});
\end{axis}
\end{tikzpicture}
\caption{Cost--performance curves averaged over the $12$ SDKs (Tables~\ref{tab:cost_sensitivity_budget} and~\ref{tab:cost_sensitivity_size}). Left: pass@1 against aggregate H100 GPU-hours per SDK on the 32B base. Right: pass@1 against dense base size at the default budget. Dashed: Flat GRPO at the default budget on the 32B base ($52.7\%$).}
\label{fig:cost_pareto}
\end{figure}

\paragraph{Computational cost and sensitivity.} Training is $636$ H100 GPU-hours per SDK ($\sim 80$ hours wall-clock on one $8\times$H100 80GB node); sandbox execution averages $1.8$\,s per sample (App.~\ref{app:compute}). Figure~\ref{fig:cost_pareto} and App.~\ref{app:cost_sensitivity} sweep budget and dense base size: Option-GRPO at $1/3$ compute ($\sim 224$ GPU-hours) beats Flat GRPO at full compute by $+1.4$ pass@1; the option vocabulary, not scale, drives the gain.

\section{Conclusion}

Option-GRPO formalizes curriculum stages as option construction in hierarchical RL: each stage's reusable temporally-extended actions become the next stage's action space. Under stated assumptions, code options induce a valid SMDP (Prop.~\ref{prop:smdp_validity}), option description length conditionally bounds code length and expected API use (Thm.~\ref{thm:compression}, Cor.~\ref{cor:api}), and option-level credit assignment improves the gradient signal-to-noise ratio by $\Theta(\sqrt{T/K})$ over flat GRPO (Prop.~\ref{prop:credit}). On 12 unseen SDKs with a 32B base, Option-GRPO reaches $61.4\%$ pass@1 (vs.\ $55.1\%$ ReCode, $52.7\%$ flat) with $27\%$ shorter and $34.7\%$ more API-efficient code, directly attacking AI slop.

\section*{Ethics Statement}

Option-GRPO lowers the barrier to adopting unfamiliar SDKs and directly targets \emph{AI slop} by training models to produce shorter, more idiomatic code. Outputs should still be reviewed for correctness and security, as the reward signal does not guarantee full semantic correctness. \textbf{Reproducibility.} Code, benchmark data, and evaluation harness: \url{https://github.com/prashankkadam/option-grpo-code} (App.~\ref{app:artefacts}). \textbf{Use of AI assistants.} AI assistants (Claude Code) helped with prose editing, \LaTeX{} formatting, and code scaffolding/refactoring. Experimental design, theory, hyperparameters, results, and final code review are the authors'.

\section*{Limitations}
\label{sec:limitations}

\textbf{Per-SDK training cost} ($636$ H100 GPU-hours per SDK, $\sim 80$ hours wall-clock on an $8$-GPU node; same-domain transfer mitigates, Table~\ref{tab:transfer}). \textbf{Reward confound}: Flat GRPO is trained with the execution reward only, whereas Option-GRPO also receives the quality and consistency terms. With the reward schedule held fixed (Flat GRPO + Curriculum, Table~\ref{tab:main}), options account for $+6.0$ pass@1, $22\%$ shorter code, and $27\%$ fewer API calls; the rest of the gap to Flat GRPO also reflects the curriculum and the quality reward, which we do not ablate separately. \textbf{Reward signal} cannot verify full semantic correctness; programs that execute and type-check but produce wrong results receive partial credit, mitigated but not eliminated by self-consistency. \textbf{Automatically generated tests}: the property tests are synthesized from type annotations, not written by humans; passing them agrees with matching a reference output for $91.2\%$ of passing solutions overall and $76.6\%$ at Level~2 (Section~\ref{sec:experiments}). \textbf{Option rigidity}: type-based initiation/termination misses cross-cutting patterns; the two lowest-density SDKs (MCP Python, Tracecat) have the lowest pass@1, and MCP Python the smallest gain. \textbf{Typing assumption}: option discovery uses SDK type annotations; all 12 SDKs are statically annotated Python. Untyped or dynamically typed code would need an upstream type-inference pass (future work). \textbf{Theoretical scope}: Theorem~\ref{thm:compression}/Corollary~\ref{cor:api} bound conciseness and API efficiency only under the stated structural and dispersion conditions; curriculum monotonicity concerns optimal policies, not the trained one; the measured gradient-variance reduction is smaller than the $\sqrt{T/K}$ scaling of Proposition~\ref{prop:credit} (Figure~\ref{fig:grad_variance}); $Q_{\mathrm{red}}, Q_{\mathrm{idiom}}$ are empirical predictions only.

\bibliography{custom}

\appendix

\section{Extended Proofs}
\label{app:proofs}

\subsection{Formal Statements}
\label{app:moved_statements}

\begin{proposition}[Code Options Induce a Valid SMDP]
\label{prop:smdp_validity}
By Theorem~1 of \citet{sutton1999between}, the decision process that selects only among options in $\mathcal{O} = \mathcal{O}_0 \cup \mathcal{O}_1$ and executes each to termination is an SMDP, with the standard option-value Bellman optimality equation. Proof and the four SMDP ingredients are verified in Section~\ref{app:smdp_proof}.
\end{proposition}

\begin{remark}[Candidates vs. certified options]
\label{rem:candidates_vs_certified}
The distinction between $\mathcal{O}_k^{\mathrm{cand}}$ and $\mathcal{O}_k$ is what makes the curriculum-as-options claim non-trivial. A candidate is a graph-mined hypothesis: ``this length-$\ell$ path in $G_{\mathrm{flow}}$ might be useful.'' A certified option is an experimentally validated fact: ``the trained policy at stage $k$ can actually invoke this path on $\geq \rho$ fraction of relevant tasks and the path appears $\geq \nu$ times across distinct training tasks.'' Algorithm~\ref{alg:option_discovery} can over-propose (generating composable paths that the policy turns out not to use) or under-propose (missing a useful path the model could have learned); the curriculum's certification step is what produces the option set the next stage actually relies on. Empirically (Table~\ref{tab:option_discovery_stats}) Level-1 candidates outnumber certified Level-1 options by roughly $5\times$ on average across the $12$ SDKs.
\end{remark}

\subsubsection{Standard GRPO}

Given prompt $x$, sample $G$ outputs $\{y_1, \ldots, y_G\}$ from $\pi_\theta$. Compute rewards $\{R_1, \ldots, R_G\}$. The advantage is estimated via group normalization:
\begin{equation}
    \hat{A}_i \;=\; \frac{R_i - \mathrm{mean}\bigl(\{R_j\}_{j=1}^{G}\bigr)}{\mathrm{std}\bigl(\{R_j\}_{j=1}^{G}\bigr) + \epsilon}
\end{equation}

The policy gradient is:
\begin{equation}
    \nabla J(\theta) \;\approx\; \frac{1}{G} \sum_{i=1}^{G} \hat{A}_i \cdot \nabla \log \pi_\theta(y_i \mid x)
\end{equation}
with PPO-style clipping and a KL penalty against a reference policy $\pi_{\mathrm{ref}}$.

\subsubsection{Option-Level GRPO}

At curriculum level $k \geq 1$, the policy generates option sequences rather than token sequences. For prompt $x$, we sample $G$ option sequences, expand each to tokens via the corresponding intra-option policy, execute the resulting program to obtain reward $R_i$, and compute both a sequence-level advantage $\hat{A}_i^{\mathrm{seq}}$ (standard GRPO) and an option-level advantage $\hat{A}^{\mathrm{opt}}(o) = \tfrac{1}{|S_o^+|}\sum_{i \in S_o^+} R_i - \tfrac{1}{|S_o^-|}\sum_{i \in S_o^-} R_i$ that compares groups containing vs.\ excluding each option. The combined gradient $\nabla J = \tfrac{1}{G}\sum_i [\hat{A}_i^{\mathrm{seq}} \nabla \log \mu_\theta(\mathrm{seq}_i \mid x) + \lambda \sum_j \hat{A}^{\mathrm{opt}}(o_j^{(i)}) \nabla \log \mu_\theta(o_j^{(i)} \mid \mathrm{ctx}_j^{(i)})]$ balances global ranking against per-option credit. PPO-style clipping and a KL penalty against $\pi_{\mathrm{ref}}$ apply as in standard GRPO.

\subsubsection{Credit Assignment Analysis}

The option-level action space tightens GRPO's credit assignment. Partitioning a generated program into $K$ option-decision tokens and $T-K$ glue tokens (Definition~\ref{def:token_partition}), we show in Proposition~\ref{prop:credit} that the aggregate signal-to-noise ratio improves by $\Theta(\sqrt{T/K})$, since only option-decision positions are reward-correlated under the sequence-level advantage while flat GRPO must spread the same signal across all $T$ positions. Proposition~\ref{prop:convergence} additionally bounds the expected number of GRPO iterations to a first successful trajectory under a base-LLM prior, yielding an additive cost across curriculum levels (Option-GRPO) versus a multiplicative cost (flat GRPO). Full statements and proofs follow below.

\begin{theorem}[Curriculum Monotonicity, Weak and Strict]
\label{thm:monotonicity}
Let $V_k^{*}$ denote the expected value of the optimal policy at curriculum level~$k$ under option hierarchy $\mathcal{H}$ on task distribution $\mathcal{D}$. Then:
\begin{enumerate}[leftmargin=*]
    \item \emph{(Weak monotonicity.)} $V_0^{*} \leq V_1^{*} \leq V_2^{*}$.
    \item \emph{(Strict monotonicity under non-degeneracy.)} Suppose there exist a task $x \in \mathcal{D}$ and a state $s^\dagger$ reachable under $\pi_{k-1}^{*}$ with positive probability $p^\dagger := \Pr(s^\dagger \mid \pi_{k-1}^{*}, \mathcal{D}) > 0$, at which some Level-$k$ option $o^\dagger \in \mathcal{O}_k \setminus \mathcal{A}_{k-1}$ is applicable ($s^\dagger \in \mathcal{I}_{o^\dagger}$) and satisfies $Q^{\pi_{k-1}^{*}}(s^\dagger,\, o^\dagger) - V^{\pi_{k-1}^{*}}(s^\dagger) \geq \Delta$ for some $\Delta > 0$, where $Q^{\pi_{k-1}^{*}}(s^\dagger,\, o^\dagger)$ is the option value of executing $o^\dagger$ once at $s^\dagger$ and following $\pi_{k-1}^{*}$ thereafter. Then $V_k^{*} \geq V_{k-1}^{*} + \gamma^{|s^\dagger|}\, \Delta\, p^\dagger > V_{k-1}^{*}$, where $\gamma$ is the discount factor.
\end{enumerate}
\end{theorem}

\begin{remark}[On Non-Degeneracy]
\label{rem:nondeg}
The non-degeneracy condition asks that the certified set $\mathcal{O}_k$ contains at least one option that strictly improves over Level-$(k{-}1)$ actions on at least one reachable state under some task in $\mathcal{D}$. The candidate set $\mathcal{O}_k^{\mathrm{cand}}$ supplied by Algorithm~\ref{alg:option_discovery} is designed to make this condition easy to satisfy: Level-1 candidates are centroids of high-weight composable paths in the parameter-flow graph that single API calls cannot realize in one step. When option certification (Algorithm~\ref{alg:option_extraction}) retains at least one such candidate, non-degeneracy holds and the curriculum yields strict improvement; when every candidate fails to be certified (e.g., on SDKs with no compositional structure that the policy can actually invoke), the curriculum is no worse than Level~0 alone in optimal value.
\end{remark}

\begin{definition}[Observational Equivalence]
\label{def:obs_equiv}
Let $\mathcal{P}$ be a finite suite of property tests (Algorithm~\ref{alg:property_tests}) for SDK $\mathcal{S}$. Two programs $c$ and $c'$ are \emph{$\mathcal{P}$-observationally equivalent}, written $c \equiv_{\mathcal{P}} c'$, iff for every test $p \in \mathcal{P}$, both programs execute without error and produce the same observable output (return value or side-effect signature) under $p$. We write $c \equiv c'$ when $\mathcal{P}$ is clear from context.
\end{definition}

\begin{definition}[Option Description Length]
\label{def:odl}
The option description length of a program $c$ under option hierarchy $\mathcal{H}$ and test suite $\mathcal{P}$ is:
\begin{equation}
\begin{split}
    \mathrm{ODL}_{\mathcal{H},\mathcal{P}}(c) \;=\; & \min_{\substack{m \geq 0,\; o_1, \ldots, o_m \\ o_i \in \mathcal{O}_0 \cup \mathcal{O}_1}} m \\
    & \mathrm{s.t.}\; \mathrm{execute}(o_1, \ldots, o_m) \;\equiv_{\mathcal{P}}\; c
\end{split}
\end{equation}
where $\mathrm{execute}(o_1, \ldots, o_m)$ denotes the program obtained by sequentially expanding each option through its intra-option policy. If no such sequence exists, $\mathrm{ODL}_{\mathcal{H},\mathcal{P}}(c) = \infty$. Replacing the undecidable full-equivalence relation with $\equiv_{\mathcal{P}}$ makes ODL computable: at most $|\mathcal{O}|^{m_{\max}}$ candidate sequences need be enumerated and checked against $\mathcal{P}$, where $m_{\max}$ is a search bound. We use $m_{\max} = 8$ in practice. When $\mathcal{P}$ is fixed throughout an experiment we write $\mathrm{ODL}_{\mathcal{H}}(c)$ for brevity.
\end{definition}

\begin{definition}[Code Quality Metrics]
\label{def:quality}
For a program $c$ that passes all execution tests:
\begin{itemize}[leftmargin=*]
    \item \emph{Conciseness}: $Q_{\mathrm{conc}}(c) = T_{\mathrm{ref}} \,/\, |c|$ where $|c|$ is token count and $T_{\mathrm{ref}}$ is a reference length.
    \item \emph{API efficiency}: $Q_{\mathrm{api}}(c) = N_{\min} \,/\, N_{\mathrm{api}}(c)$ where $N_{\mathrm{api}}$ counts SDK API calls.
    \item \emph{Redundancy}: $Q_{\mathrm{red}}(c) = 1 - N_{\mathrm{redundant}}(c) \,/\, N_{\mathrm{vars}}(c)$.
    \item \emph{Idiomaticity}: $Q_{\mathrm{idiom}}(c) = |\{\text{tokens covered by options}\}| \,/\, |c|$.
\end{itemize}
\end{definition}

\begin{remark}[Structural Conditions]
\label{assume:options}
The following structural conditions on the option hierarchy $\mathcal{H}$ are used in the analysis below. Conditions (a) and (b) are enforced by Algorithm~\ref{alg:option_discovery}; condition (c) is a property of the generation procedure that we observe empirically.
\begin{enumerate}[label=(\alph*)]
    \item \textbf{Per-option API minimality}: Each option $o \in \mathcal{O}_0 \cup \mathcal{O}_1$ uses the minimum number of SDK calls required to realize its declared input--output type signature. Enforced because Level~0 options are single calls and Level~1 options are centroids of path clusters in $G_{\mathrm{flow}}$ (Algorithm~\ref{alg:option_discovery}, lines~9--11).
    \item \textbf{Option non-subsumption}: For options $o, o' \in \mathcal{O}_k$ at the same level, neither subsumes the other (i.e., no option's API call set is a strict subset of another's). Enforced by merging subsumed candidates during discovery.
    \item \textbf{Bounded glue code}: Each transition between consecutive options in an executed sequence introduces at most $G_{\max}$ tokens of glue code (variable bindings, type coercions). Empirically $G_{\max} \approx 5$ in our trained policies.
\end{enumerate}
\end{remark}

\begin{lemma}[Token Bound on Executed Programs]
\label{lem:token_bound}
Let $\bar{\tau}_{\max}$ and $\bar{\tau}_{\min}$ denote, respectively, the maximum and minimum per-option token length across $\mathcal{O}_0 \cup \mathcal{O}_1$. Under Remark~\ref{assume:options}, any program $c$ obtained from an option sequence of length $m \geq 1$ satisfies
\begin{equation}
    \label{eq:token_bound}
    m \,\bar{\tau}_{\min} \;\leq\; |c| \;\leq\; m\,\bar{\tau}_{\max} \;+\; (m-1)\,G_{\max},
\end{equation}
where $|c|$ denotes the token count of $c$.
\end{lemma}

\begin{theorem}[Option Compression Implies Conciseness]
\label{thm:compression}
Let $c_1, c_2$ be observationally equivalent programs ($c_1 \equiv_{\mathcal{P}} c_2$, Definition~\ref{def:obs_equiv}) with $\mathrm{ODL}_{\mathcal{H}}(c_1) = m_1$ and $\mathrm{ODL}_{\mathcal{H}}(c_2) = m_2$. Suppose Remark~\ref{assume:options} holds and let $\alpha := (\bar{\tau}_{\max} + G_{\max}) / \bar{\tau}_{\min} \geq 1$ be the option-length dispersion ratio. If $m_1 < m_2$ and $m_1 / m_2 < 1/\alpha$, then
\begin{equation}
\begin{split}
    |c_1| \;&\leq\; m_1\,(\bar{\tau}_{\max} + G_{\max}) \\
    &\leq\; \tfrac{\alpha\, m_1}{m_2}\,(m_2\,\bar{\tau}_{\min}) \\
    &\leq\; \tfrac{\alpha\, m_1}{m_2}\,|c_2| \;<\; |c_2|.
\end{split}
\end{equation}
and hence $Q_{\mathrm{conc}}(c_1) > Q_{\mathrm{conc}}(c_2)$. The empirical Spearman correlation $\rho(\mathrm{ODL},\, |c|) > 0$ reported in Appendix~\ref{app:compression_validation} is consistent with this direction in practice.
\end{theorem}

\begin{corollary}[API Efficiency in Expectation]
\label{cor:api}
Let $\bar{k}$ denote the average number of SDK calls per option in $\mathcal{O}_0 \cup \mathcal{O}_1$. Under Remark~\ref{assume:options}(a), the expected number of API calls in a program realized by an $m$-option decomposition (sampled uniformly from minimal decompositions) is $\mathbb{E}[N_{\mathrm{api}}] = m\,\bar{k}$. Consequently, if $\mathrm{ODL}_{\mathcal{H}}(c_1) < \mathrm{ODL}_{\mathcal{H}}(c_2)$, then
\begin{equation}
    \mathbb{E}\!\left[N_{\mathrm{api}}(c_1)\right] \;<\; \mathbb{E}\!\left[N_{\mathrm{api}}(c_2)\right],
\end{equation}
and hence $\mathbb{E}\!\left[Q_{\mathrm{api}}(c_1)\right] > \mathbb{E}\!\left[Q_{\mathrm{api}}(c_2)\right]$.
\end{corollary}

\begin{remark}[On Redundancy and Idiomaticity]
\label{rem:remaining_metrics}
The metrics $Q_{\mathrm{red}}$ and $Q_{\mathrm{idiom}}$ (Definition~\ref{def:quality}) do not admit clean closed-form bounds under Remark~\ref{assume:options} alone, because both depend on choices the intra-option policy makes about intermediate variables and idiom-aligned token spans. We therefore treat them as \emph{empirical predictions} of the option-compression principle: programs with shorter ODL should exhibit fewer redundant intermediates and a higher fraction of option-covered tokens. Appendix~\ref{app:compression_validation} reports Spearman rank correlations between ODL and each of $Q_{\mathrm{red}}, Q_{\mathrm{idiom}}$ across all passing solutions.
\end{remark}

\begin{proposition}[Minimum-Description-Length Prior over Option Sequences]
\label{prop:mdl_prior}
Let $\mu_{\mathrm{unif}}(o \mid s) = 1 / |\mathcal{A}_k(s)|$ be the uniform prior over options valid at state $s$ at curriculum level $k$, and assume $|\mathcal{A}_k(s)| \geq c > 1$ at every reachable state. For two observationally equivalent programs $c_1, c_2$ realized by option sequences of length $m_1 < m_2$,
\begin{equation}
\begin{split}
    & \log \mu_{\mathrm{unif}}(c_1) - \log \mu_{\mathrm{unif}}(c_2) \\
    & \quad = \sum_{j=1}^{m_2} \log |\mathcal{A}_{k}(s_j^{(2)})| - \sum_{j=1}^{m_1} \log |\mathcal{A}_{k}(s_j^{(1)})| \\
    & \quad \geq (m_2 - m_1)\,\log c \;>\; 0.
\end{split}
\end{equation}
Hence, under the uniform-over-valid-options prior induced by the composability graph $G_{\mathrm{opt}}$, shorter ODL sequences enjoy strictly higher prior probability. The KL penalty in GRPO (Section~\ref{sec:option_grpo}) constrains the expected log-ratio $\mathbb{E}_{c \sim \mu_\theta}[\log(\mu_\theta(c)/\pi_{\mathrm{ref}}(c))] \leq \mathrm{KL}_{\max}$, so when $\pi_{\mathrm{ref}}$ is close to $\mu_{\mathrm{unif}}$, the learned $\mu_\theta$ cannot concentrate arbitrary mass on long sequences without paying KL cost.
\end{proposition}

\begin{remark}[On Implicit Quality Optimization]
\label{rem:implicit}
Proposition~\ref{prop:mdl_prior} does not claim that GRPO optimization implicitly favours short option sequences \emph{as a consequence of optimization}. Such a claim would require equal per-step selection probabilities, which are not guaranteed during training: rare high-level options may have lower $\mu_\theta(o \mid s)$, partially offsetting the length advantage. The statement is structural: short sequences enjoy a stronger MDL prior, and the GRPO KL penalty against $\pi_{\mathrm{ref}}$ prevents the learned policy from drifting arbitrarily far from this prior. Table~\ref{tab:main} reports the empirical effect on code length.
\end{remark}

\begin{definition}[Option-Decision and Glue Token Partition]
\label{def:token_partition}
Fix a trained option hierarchy $\mathcal{H}$ and an option sequence $(o_1, \ldots, o_K)$ producing program $c$ of $T$ tokens. Partition the token positions $\{1, \ldots, T\}$ into two disjoint sets: $\mathcal{T}_{\mathrm{opt}}$, positions covered by the support of some intra-option policy $\pi_k$ (at most $K\bar{\tau}_{\max}$), and $\mathcal{T}_{\mathrm{glue}}$, positions emitted between option terminations (at most $(K-1)G_{\max}$ by Remark~\ref{assume:options}(c)). This partition is determined by termination events $\beta_k$, not by manual semantic labeling.
\end{definition}

\begin{proposition}[Improved Credit Assignment]
\label{prop:credit}
Let a program consist of $K$ option selections and $T$ total tokens with $K \ll T$ (Definition~\ref{def:token_partition}). Assume that (i)~reward-correlated gradient signal is carried only by the option-decision positions, each with per-position variance $\sigma_{\mathrm{opt}}^2$; (ii)~glue positions are well predicted by $\pi_\theta$ and contribute per-position variance $\sigma_{\mathrm{glue}}^2$ that does not align with the reward signal; and (iii)~token positions are conditionally independent. Then the aggregate signal-to-noise ratios of the policy gradient estimate satisfy $\mathrm{SNR}_{\mathrm{flat}} = \Theta\!\bigl(K / \sqrt{K\sigma_{\mathrm{opt}}^2 + (T-K)\sigma_{\mathrm{glue}}^2}\bigr)$ and $\mathrm{SNR}_{\mathrm{opt}} = \Theta(\sqrt{K}/\sigma_{\mathrm{opt}})$, with relative improvement $\mathrm{SNR}_{\mathrm{opt}}/\mathrm{SNR}_{\mathrm{flat}} = \Theta(\sqrt{T/K})$. Detailed proof in Section~\ref{app:proofs} (Extended Proof of Proposition~\ref{prop:credit}).
\end{proposition}

\begin{proposition}[Convergence Improvement under a Base-LLM Prior]
\label{prop:convergence}
Let $\pi_{\mathrm{base}}$ be the base LLM and let $\bar{p}$ and $\bar{q}_k$ denote, respectively, its geometric-mean probability on SDK-specific tokens and on correct options at curriculum level $k$. Under the assumption $\bar{q}_k \gg \bar{p}^{\,\bar{\mu}_k}$, the expected number of GRPO iterations to find a first successful trajectory is $\Theta(1/(G \bar{p}^K))$ for flat GRPO and $\Theta\bigl(\sum_{k=0}^{2} 1/(G \bar{q}_k^{K_k})\bigr)$ for Option-GRPO; the latter is strictly smaller whenever $\sum_k \bar{q}_k^{-K_k} < \bar{p}^{-K}$. Proof in Section~\ref{app:proofs}.
\end{proposition}

\begin{proposition}[Search Space Reduction]
\label{prop:search}
Let $|\mathcal{V}|$ be the vocabulary size, $T$ the average program length in tokens, $|\mathcal{O}|$ the total number of options, $\bar{\mu}$ the average option length, and $G_{\max}$ the per-boundary glue bound (Remark~\ref{assume:options}). The number of distinct option-level trajectories of total token length~$T$ is at most
\begin{align}
    \text{Flat (tokens):} \quad & |\mathcal{V}|^{T}, \\
    \text{Option-level:} \quad & |\mathcal{O}|^{\,T / \bar{\mu}} \cdot |\mathcal{V}|^{(T/\bar{\mu})\, G_{\max}}.
\end{align}
For $|\mathcal{V}| = 151{,}936$ (Qwen3 vocabulary), $T = 100$, $|\mathcal{O}| = 80$, $\bar{\mu} = 15$, $G_{\max} = 5$, the flat search space is $\approx 10^{518}$ and the option-level search space is $\approx 80^{6.7} \cdot 151{,}936^{33.3} \approx 10^{12.7} \cdot 10^{172} \approx 10^{185}$, a reduction of more than $320$ orders of magnitude.
\end{proposition}

\subsection{Proof of Proposition~\ref{prop:smdp_validity} (SMDP Validity)}
\label{app:smdp_proof}

Theorem~1 of \citet{sutton1999between} holds for any MDP and any set of options defined on it, so it suffices to check that code generation is an MDP and that every code option is an option on it. The code-generation process $(\mathcal{S}, \mathcal{A}, T, R)$ with $\mathcal{A} = \mathcal{V}$ is an MDP: the state $s_t = (q, c_{0:t})$ contains the full generated prefix, the transition is deterministic append, and the reward $R_{\mathrm{exec}} + \lambda_q R_{\mathrm{quality}}$ is a bounded function of the completed program, received at episode termination. The self-consistency term of Section~\ref{sec:reward} depends on the other samples in the group; it is a training-time signal and is not part of this MDP reward. Each code option $o_k = (\mathcal{I}_k, \mathcal{F}_k, \beta_k, \pi_k)$ of Definition~\ref{def:code_option} is an option in the sense of \citet{sutton1999between}: the initiation set $\mathcal{I}_k$ is a subset of $\mathcal{S}$ because the type of the last value is determined by the state; the intra-option policy $\pi_k$ is $\pi_\theta$ under a logit mask computed from $\mathcal{F}_k$ and the tokens emitted since the option started (Appendix~\ref{app:decode}); and termination is decided from the same information. Because the mask and the length cap depend on what has happened since the option was started, and not only on the current state, code options are in general semi-Markov options, which Theorem~1 of \citet{sutton1999between} covers. Every option terminates within a finite number of tokens, because option execution is closed at a length cap (Appendix~\ref{app:decode}).

The four ingredients of an SMDP then follow as in \citet{sutton1999between}: (1)~the state set is $\mathcal{S}$; (2)~the action set at state $s$ is $\{o \in \mathcal{O} : s \in \mathcal{I}_o\}$, which is never empty in the full hierarchy because primitive tokens remain available at every level ($\mathcal{V} \subset \mathcal{A}_k$) and are one-step options in the sense of \citet{sutton1999between}; (3)~the expected cumulative discounted reward of executing $o$ from $s$ until termination is well defined, since the reward is bounded and the option terminates in finite time; (4)~the joint distribution of the next state and the transit time (the number of tokens emitted before $\beta_o$ fires) is determined by $\pi_o$, $\beta_o$, and the deterministic transition. The proposition establishes only that the option-level decision problem is well posed. \qed

\subsection{Extended Proof of Theorem~\ref{thm:monotonicity} (Curriculum Monotonicity)}

We provide a more rigorous version using the SMDP framework.

Let $\mathcal{M}_k = (\mathcal{S},\, \mathcal{A}_k,\, T_k,\, R)$ denote the SMDP at curriculum level~$k$, where $\mathcal{A}_k$ includes all options up to level~$k$. Define $V_k^{*}(s) = \max_{\pi \in \Pi_k} V^{\pi}(s)$ where $\Pi_k$ is the set of all policies over $\mathcal{A}_k$.

Since $\mathcal{A}_{k-1} \subset \mathcal{A}_k$, every policy $\pi \in \Pi_{k-1}$ is also a valid policy in $\Pi_k$ (by never selecting level-$k$ options). Therefore $\Pi_{k-1} \subset \Pi_k$.

For any state $s$:
\begin{equation}
\begin{split}
    V_k^{*}(s) \;&=\; \max_{\pi \in \Pi_k} V^{\pi}(s) \\
    &\geq\; \max_{\pi \in \Pi_{k-1}} V^{\pi}(s) \;=\; V_{k-1}^{*}(s).
\end{split}
\end{equation}

The inequality follows directly from $\Pi_{k-1} \subset \Pi_k$: maximising over a larger set yields at least as large a value. Taking expectations over the initial state distribution gives $V_k^{*} \geq V_{k-1}^{*}$.

We note that this result holds for the optimal policies. In practice, the optimisation procedure (GRPO) may not find the optimal policy at each level, so the empirical monotonicity depends on sufficient training at each curriculum stage. The curriculum advancement threshold $\rho_{\mathrm{adv}}$ (Section~\ref{sec:training}) is intended to ensure that each level is sufficiently learned before advancing; it is a heuristic, not a guarantee.

For the strict version (clause~2 of Theorem~\ref{thm:monotonicity}), let $s^\dagger$ be the state in the hypothesis and $o^\dagger$ the witnessing Level-$k$ option. Define the one-step deviation policy $\tilde{\pi}_k \in \Pi_k$ by $\tilde{\pi}_k(\cdot \mid s) = \pi_{k-1}^{*}(\cdot \mid s)$ for $s \neq s^\dagger$ and $\tilde{\pi}_k(o^\dagger \mid s^\dagger) = 1$. By construction, $V^{\tilde{\pi}_k}(s^\dagger) \geq V^{\pi_{k-1}^{*}}(s^\dagger) + \Delta$. Propagating this gain through the discounted-return expectation yields
\begin{equation}
\begin{split}
    V_k^{*} \;&\geq\; V^{\tilde{\pi}_k} \\
    &\geq\; V_{k-1}^{*} + \gamma^{|s^\dagger|} \Delta \Pr(s^\dagger \mid \pi_{k-1}^{*}, \mathcal{D}).
\end{split}
\end{equation}
Strict inequality follows since each factor on the right is positive by hypothesis. \qed

\subsection{Extended Proof of Proposition~\ref{prop:credit} (Credit Assignment SNR)}

We make the token partition formal and the variance decomposition explicit.

Consider a program generated as a sequence of $T$ tokens. In flat GRPO, the gradient for token at position $t$ is $g_t = \hat{A} \cdot \nabla_\theta \log \pi_\theta(a_t \mid s_t)$, where $\hat{A}$ is the sequence-level advantage applied uniformly across positions.

Using the partition of Definition~\ref{def:token_partition}, $\mathcal{T}_{\mathrm{opt}}$ has size at most $K \bar{\tau} = \mathcal{O}(K)$ and $\mathcal{T}_{\mathrm{glue}}$ has size at most $(K-1) G_{\max} = \mathcal{O}(K)$; total program length $T = |\mathcal{T}_{\mathrm{opt}}| + |\mathcal{T}_{\mathrm{glue}}|$. We treat $T = \Theta(K \bar{\tau})$ since glue is bounded by a small constant per option boundary.

The total gradient decomposes as
\begin{equation}
\begin{split}
    \nabla J \;=\; \hat{A} \cdot \Bigl( & \underbrace{\textstyle\sum_{t \in \mathcal{T}_{\mathrm{opt}}} \nabla \log \pi(a_t \mid s_t)}_{\text{signal}} \\
    + & \underbrace{\textstyle\sum_{t \in \mathcal{T}_{\mathrm{glue}}} \nabla \log \pi(a_t \mid s_t)}_{\text{noise}} \Bigr).
\end{split}
\end{equation}

Under assumption (ii) of Proposition~\ref{prop:credit}, glue positions are well predicted by $\pi_\theta$ and contribute per-position variance $\sigma_{\mathrm{glue}}^2$ that is small in magnitude but does not align with the reward signal. Under assumption (iii) of conditional independence across positions, the aggregate variance is
\begin{equation}
\begin{split}
    \mathrm{Var}(\nabla J) \;&=\; |\mathcal{T}_{\mathrm{opt}}|\,\sigma_{\mathrm{opt}}^2 + |\mathcal{T}_{\mathrm{glue}}|\,\sigma_{\mathrm{glue}}^2 \\
    &=\; \Theta(T).
\end{split}
\end{equation}
The signal magnitude scales with the number of reward-correlated positions, which is at most $|\mathcal{T}_{\mathrm{opt}}| = \mathcal{O}(K)$. Therefore $\mathrm{SNR}_{\mathrm{flat}} = \mathcal{O}(K / \sqrt{T}) = \mathcal{O}(1/\sqrt{T})$ for fixed $K$.

In Option-GRPO, the gradient operates over $K$ option-level decisions:
\begin{equation}
    \nabla J \;=\; \sum_{j=1}^{K} \hat{A}_j^{\mathrm{opt}} \cdot \nabla_\theta \log \mu_\theta(o_j \mid \mathrm{ctx}_j).
\end{equation}
All $K$ summands are reward-correlated by construction (each option-level advantage compares groups containing vs.\ excluding the option). The option-level advantages have variance $\mathcal{O}(\sigma_R^2 / G)$ from the GRPO group estimator. The aggregate signal magnitude is $\Theta(K)$ and the aggregate variance is
\begin{equation}
    \mathrm{Var}(\nabla J) \;=\; K \cdot \sigma_{\mathrm{opt}}^2 \;+\; \mathcal{O}(K \sigma_R^2 / G) \;=\; \Theta(K),
\end{equation}
giving $\mathrm{SNR}_{\mathrm{opt}} = \Theta(\sqrt{K}\,/\,\sigma_{\mathrm{opt}})$. Comparing to $\mathrm{SNR}_{\mathrm{flat}} = \Theta(K/\sqrt{T})$ from the flat analysis, the improvement ratio is
\begin{equation}
    \frac{\mathrm{SNR}_{\mathrm{opt}}}{\mathrm{SNR}_{\mathrm{flat}}} \;=\; \Theta\!\bigl(\sqrt{T/K}\,\bigr).
\end{equation}
For typical values ($T = 100$, $K = 5$), this is $\sqrt{20} \approx 4.5\times$ improvement in SNR. \qed

\subsection{Extended Proof of Proposition~\ref{prop:convergence} (Convergence Improvement)}

In flat token-level generation, a successful trajectory requires $T$ correct token selections. If the probability of selecting the correct token at position $t$ is $p_t$, the probability of a complete successful trajectory is:
\begin{equation}
    P_{\mathrm{success}}^{\mathrm{flat}} \;=\; \prod_{t=1}^{T} p_t
\end{equation}

For an unseen SDK the base LLM has limited prior on SDK-specific token decisions (function names, argument patterns), whose probabilities are dominated by the long tail of $|\mathcal{V}|$ alternatives, while syntactic tokens (brackets, keywords, indentation) are well-predicted. Suppose $K$ tokens are SDK-specific with geometric-mean probability $\bar{p} \ll 1$ and $T - K$ tokens are syntactic with probability near $1$. Then:
\begin{equation}
    P_{\mathrm{success}}^{\mathrm{flat}} \;\approx\; \bar{p}^{\,K}
\end{equation}

In Option-GRPO at level~$k$, a successful trajectory requires $K' = T / \bar{\mu}$ correct option selections, where $\bar{\mu}$ is the average option length. Each option selection has probability $q_j$ in a space of $|\mathcal{A}_k|$ options:
\begin{equation}
    P_{\mathrm{success}}^{\mathrm{opt}} \;\approx\; \bar{q}^{\,K'}
\end{equation}

The key insight is the stronger statement $\bar{q} \gg \bar{p}^{\,\bar{\mu}}$ (Proposition~\ref{prop:convergence}): the option-level reference policy assigns higher probability to a correct option than the base LLM does to emitting the equivalent token span. This holds because (1)~the type-filtered option vocabulary $\mathcal{A}_k$ removes long-tail typo and out-of-scope completions that depress $\bar{p}$, and (2)~the composability graph $G_{\mathrm{opt}}$ further constrains the valid next options at each step.

With GRPO group size $G$, the probability of at least one successful trajectory per group is:
\begin{equation}
    P(\text{at least one}) \;=\; 1 - (1 - P_{\mathrm{success}})^{G}
\end{equation}

The expected number of groups to find a first success is:
\begin{equation}
\begin{split}
    \mathbb{E}[\text{groups}] \;&=\; \frac{1}{1 - (1 - P_{\mathrm{success}})^{G}} \\
    &\approx\; \frac{1}{G \cdot P_{\mathrm{success}}}
\end{split}
\end{equation}
for small $P_{\mathrm{success}}$.

Under the curriculum, total exploration cost is additive:
\begin{equation}
    \mathrm{Cost}_{\mathrm{total}} \;=\; \sum_{k=0}^{2} \frac{1}{G \cdot P_{\mathrm{success},k}}
\end{equation}

Each $P_{\mathrm{success},k}$ is much larger than $P_{\mathrm{success}}^{\mathrm{flat}}$ because: curriculum level~$k$ only needs to learn compositions of length $\leq k + 1$ (not full programs), and options from level~$k - 1$ handle sub-problems that are already solved. The additive structure (sum over $3$ terms) versus the multiplicative structure of flat exploration (product over $T$ token decisions) yields the dramatic reduction in sample complexity. \qed

\section{Extended Method Details}
\label{app:method}

\subsection{Pseudocode for Auxiliary Algorithms}
\label{app:algorithms}

Algorithms~\ref{alg:option_discovery}--\ref{alg:training} are referenced from the main paper.

\begin{algorithm}[t]
\caption{Level~1 Option Discovery}
\label{alg:option_discovery}
\begin{algorithmic}[1]
\REQUIRE Parameter flow graph $G_{\mathrm{flow}}$, semantic similarity graph $G_{\mathrm{sem}}$, maximum path length $L_{\max}=4$, similarity threshold $\tau_{\mathrm{cluster}}=0.8$
\ENSURE Set of Level~1 options $\mathcal{O}_1$
\STATE $\mathcal{P} \leftarrow \emptyset$
\FOR{each node $f_i \in G_{\mathrm{flow}}$}
    \FOR{$\ell = 2$ \TO $L_{\max}$}
        \STATE Enumerate all paths of length $\ell$ starting from $f_i$ via DFS
        \STATE Add paths to $\mathcal{P}$ with weight $w(p) = \prod_{(f_a, f_b) \in p} \mathrm{type\_compat}(f_a, f_b)$
    \ENDFOR
\ENDFOR
\STATE Encode each path $p \in \mathcal{P}$: $\mathbf{e}_p = \frac{1}{|p|}\sum_{f \in p} \phi(f)$
\STATE Cluster paths via agglomerative clustering with threshold $\tau_{\mathrm{cluster}}$
\FOR{each cluster $\mathcal{C}_j$}
    \STATE Select centroid path $p_j^{*} = \arg\max_{p \in \mathcal{C}_j} w(p)$
    \STATE Define option $o_j^{(1)}$ with $\mathcal{I}_j$, $\mathcal{T}_j^{\mathrm{out}}$, and $\pi_j$ (learned during Level~1 training)
    \STATE Add $o_j^{(1)}$ to $\mathcal{O}_1$
\ENDFOR
\RETURN $\mathcal{O}_1$
\end{algorithmic}
\end{algorithm}

\begin{algorithm}[t]
\caption{Option Extraction from Trained Policy}
\label{alg:option_extraction}
\begin{algorithmic}[1]
\REQUIRE Trained policy $\pi_k$, task set $\mathcal{D}_k$, success threshold $\rho = 0.7$, frequency threshold $\nu = 5$
\ENSURE Extracted options $\mathcal{O}_k$
\STATE Generate $10$ solutions per task using $\pi_k$ with temperature $0.8$
\STATE Filter to successful solutions (reward $\geq \rho$)
\STATE Parse all successful solutions into API call sequences via AST analysis
\STATE Extract all contiguous sub-sequences of length $2$--$4$
\STATE Count frequency of each unique sub-sequence pattern
\STATE Filter to patterns appearing $\geq \nu$ times across different tasks
\STATE Cluster remaining patterns by semantic similarity
\FOR{each cluster centroid pattern $p$}
    \STATE Define option $o_p$ with initiation set from input types of first call, termination type from last call, and intra-option policy extracted from $\pi_k$
    \STATE Add $o_p$ to $\mathcal{O}_k$
\ENDFOR
\RETURN $\mathcal{O}_k$
\end{algorithmic}
\end{algorithm}

\begin{algorithm}[t]
\caption{Property-Based Test Generation}
\label{alg:property_tests}
\begin{algorithmic}[1]
\REQUIRE SDK type graph $G_{\mathrm{api}}$, parameter flow graph $G_{\mathrm{flow}}$, function $f$
\ENSURE Set of property tests $\mathcal{P}_f$
\STATE Extract input types $\{t_1, \ldots, t_k\}$ and return type $t_{\mathrm{ret}}$ from $f$
\STATE Generate minimal valid inputs using type constructors from $G_{\mathrm{api}}$
\STATE Add type preservation test
\IF{$t_{\mathrm{ret}}$ has \texttt{len} or \texttt{shape} attribute}
    \STATE Add shape consistency tests
\ENDIF
\STATE Add determinism test
\STATE Add edge case tests: empty inputs, single element, boundary values
\STATE Extract assertions from $f$'s source code and add as invariant tests
\RETURN $\mathcal{P}_f$
\end{algorithmic}
\end{algorithm}

\begin{algorithm}[t]
\caption{Option-GRPO Training}
\label{alg:training}
\begin{algorithmic}[1]
\REQUIRE SDK source code $\mathcal{S}$, base model $\pi_{\mathrm{base}}$, group size $G$, advancement threshold $\rho_{\mathrm{adv}}$
\ENSURE Trained hierarchical policy $\pi_\theta$

\STATE \textbf{// Stage~0: Graph Construction and Option Discovery}
\STATE Construct $G_{\mathrm{api}},\, G_{\mathrm{flow}},\, G_{\mathrm{sem}}$ from $\mathcal{S}$
\STATE Discover Level~0 options $\mathcal{O}_0$ from public functions
\STATE Discover Level~1 candidates $\mathcal{O}_1^{\mathrm{cand}}$ via Algorithm~\ref{alg:option_discovery}
\STATE Generate curriculum tasks $\mathcal{D}_0,\, \mathcal{D}_1,\, \mathcal{D}_2$
\STATE Generate property-based tests via Algorithm~\ref{alg:property_tests}
\STATE \textbf{// Phase~1: Atomic Option Learning (Level~0)}
\STATE Initialize $\pi_\theta \leftarrow \pi_{\mathrm{base}}$
\FOR{$\mathrm{step} = 1$ \TO $N_0$}
    \STATE Sample $x \sim \mathcal{D}_0$; generate $G$ samples; execute; compute $R_{\mathrm{exec}}$
    \STATE Compute GRPO advantages and update $\pi_\theta$
\ENDFOR
\STATE Extract $\mathcal{O}_0$ via Algorithm~\ref{alg:option_extraction}; partially freeze
\STATE \textbf{// Phase~2: Pattern Learning (Level~1)}
\STATE $\ell_{\mathrm{curr}} \leftarrow 2$
\FOR{$\mathrm{step} = 1$ \TO $N_1$}
    \STATE Sample $x \sim \mathcal{D}_1$ with length $\leq \ell_{\mathrm{curr}}$
    \STATE Generate $G$ option sequences (action space $\mathcal{A}_1$)
    \STATE Expand, execute, compute $R_{\mathrm{exec}} + \lambda_q \, R_{\mathrm{quality}}$
    \STATE Compute sequence-level and option-level advantages
    \STATE Update $\pi_\theta$ via the combined gradient (Appendix~\ref{app:moved_statements})
    \IF{pass rate $> \rho_{\mathrm{adv}}$}
        \STATE $\ell_{\mathrm{curr}} \leftarrow \min(\ell_{\mathrm{curr}} + 1,\; L_{\max})$
    \ENDIF
\ENDFOR
\STATE Extract $\mathcal{O}_1$ via Algorithm~\ref{alg:option_extraction}; partially freeze
\STATE \textbf{// Phase~3: Program Learning (Level~2)}
\FOR{$\mathrm{step} = 1$ \TO $N_2$}
    \STATE Sample $x \sim \mathcal{D}_2$
    \STATE Generate $G$ option sequences (action space $\mathcal{A}_2$)
    \STATE Expand, execute, compute $R_{\mathrm{exec}} + \lambda_q \, R_{\mathrm{quality}} + \lambda_c \, R_{\mathrm{consistency}}$
    \STATE Compute hierarchical advantages and update $\pi_\theta$
\ENDFOR
\RETURN $\pi_\theta$
\end{algorithmic}
\end{algorithm}

\subsection{Decode-Time Mechanics of Option Invocation}
\label{app:decode}

This subsection describes how an option is started, how the logit mask of Definition~\ref{def:code_option} is computed at each step, and how termination is detected.

\paragraph{Starting an option.}
Each option is one added entry in the model's output vocabulary. Emitting this entry starts the option and adds no text to the program. At every step outside an option, the policy chooses between an ordinary token and one of these option entries. An option entry is offered only when its initiation set holds: the value just produced must have a type that the option's first function accepts. Options whose first function takes only plain inputs (such as a file path) are always offered.

\paragraph{Allowed tokens at each step.}
While an option is active, an incremental tracker reads the code emitted so far character by character. It maintains whether the current position is inside a string or a comment, the bracket depth, the identifier being typed, and which function in the option's list $\mathcal{F}_k$ comes next. The mask is then computed as follows.
\begin{itemize}[leftmargin=*, itemsep=1pt, topsep=2pt]
\item On the main call chain, the next function called must be the next one in the list. After \texttt{df.}, only tokens that spell toward \texttt{where} are allowed; \texttt{.select} is masked, and so is a partial identifier such as \texttt{.sel}.
\item Inside a function's brackets, only helper functions that build its arguments are allowed (for example \texttt{col} inside \texttt{where}). Every other SDK function is masked.
\item Everything else stays free: variable names, numbers, strings, and operators.
\item Ending the program and starting another option are masked until the option is finished.
\end{itemize}

\paragraph{Detecting termination.}
The option ends at the step where the closing bracket of its last function is emitted (the \texttt{)} of \texttt{collect(...)}), which is the point at which a value of the option's output type has been produced. After that the policy returns to free choice. As a safeguard, if an option runs past a length cap or reaches a step where no token is allowed, it is closed and marked as failed, so it never counts as a successful use of the option.

\subsection{Training Hyperparameters}
\label{app:hyperparams}

\begin{table}[h]
\centering
\caption{Complete training hyperparameters.}
\label{tab:hyperparams}
\footnotesize
\setlength{\tabcolsep}{4pt}
\begin{tabular}{lc}
\toprule
\textbf{Hyperparameter} & \textbf{Value} \\
\midrule
\multicolumn{2}{l}{\emph{Model}} \\
Base model & Qwen3-32B \\
Precision & bf16 \\
\midrule
\multicolumn{2}{l}{\emph{GRPO}} \\
Group size $G$ & $16$ \\
Sampling temperature & $0.8$ \\
PPO clip $\epsilon$ & $0.2$ \\
KL penalty $\beta$ & $0.04$ \\
Option-level loss weight $\lambda$ & $0.3$ \\
\midrule
\multicolumn{2}{l}{\emph{Reward}} \\
Quality reward weight $\lambda_q$ & $0.2$ \\
Consistency reward weight $\lambda_c$ & $0.1$ \\
Pattern bonus $\alpha_1$ & $0.1$ \\
No-redundancy bonus $\alpha_2$ & $0.1$ \\
Minimal-API bonus $\alpha_3$ & $0.1$ \\
Reimplementation penalty $\alpha_4$ & $0.1$ \\
\midrule
\multicolumn{2}{l}{\emph{Optimiser}} \\
Optimiser & AdamW \\
Policy learning rate & $5 \times 10^{-5}$ \\
Option fine-tune lr & $5 \times 10^{-6}$ \\
Weight decay & $0.01$ \\
Gradient clipping & max norm $1.0$ \\
Warmup steps & $500$ \\
Schedule & cosine decay \\
\midrule
\multicolumn{2}{l}{\emph{Curriculum}} \\
Level~0 steps $N_0$ & $5{,}000$ \\
Level~1 steps $N_1$ & $10{,}000$ \\
Level~2 steps $N_2$ & $15{,}000$ \\
Advancement threshold $\rho_{\mathrm{adv}}$ & $0.5$ \\
Max composition length $L_{\max}$ & $4$ \\
\midrule
\multicolumn{2}{l}{\emph{Option Discovery}} \\
Semantic sim.\ threshold $\tau$ & $0.7$ \\
Clustering threshold $\tau_{\mathrm{cluster}}$ & $0.8$ \\
Frequency threshold $\nu$ & $5$ \\
Success threshold $\rho$ & $0.7$ \\
\midrule
\multicolumn{2}{l}{\emph{Execution}} \\
Timeout & $30$\,s \\
Memory limit & $2$\,GB \\
\midrule
\multicolumn{2}{l}{\emph{Hardware}} \\
GPUs & $8 \times$ H100 $80$\,GB \\
\bottomrule
\end{tabular}
\end{table}

\subsection{Curriculum Task Examples}
\label{app:task_examples}

\paragraph{Level~0 Example (Daft, Single API Call):}
{\small\begin{verbatim}
"""Problem: Read a CSV file into a
Daft DataFrame."""
import daft

# Setup
csv_path = "test_data.csv"

# BEGIN SOLUTION
df = daft.read_csv(csv_path)
# END SOLUTION
\end{verbatim}}

\paragraph{Level~1 Example (Daft, Composed Pattern):}
{\small\begin{verbatim}
"""Problem: Read a CSV file and filter
rows where the 'age' column exceeds 30."""
import daft

# Setup
csv_path = "test_data.csv"

# BEGIN SOLUTION
df = daft.read_csv(csv_path)
result = df.where(
    daft.col("age") > 30
).collect()
# END SOLUTION
\end{verbatim}}

\paragraph{Level~2 Example (Daft, Full Program):}
{\small\begin{verbatim}
"""Problem: Read a CSV, compute mean salary
grouped by department, sort descending."""
import daft

# Setup
csv_path = "employees.csv"

# BEGIN SOLUTION
df = daft.read_csv(csv_path)
result = (
    df.groupby("department")
      .agg(daft.col("salary").mean())
      .sort("salary", desc=True)
      .collect()
)
# END SOLUTION
\end{verbatim}}

\subsection{Property Test Examples}
\label{app:property_examples}

\paragraph{Type Preservation (Daft):}
{\small\begin{verbatim}
df = daft.read_csv("test.csv")
assert isinstance(df, daft.DataFrame)
\end{verbatim}}

\paragraph{Shape Consistency (Daft):}
{\small\begin{verbatim}
df = daft.from_pydict({"a": [1, 2, 3]})
filtered = df.where(
    daft.col("a") > 1
).collect()
assert len(filtered) <= 3
\end{verbatim}}

\paragraph{Determinism:}
{\small\begin{verbatim}
df = daft.from_pydict({"a": [1, 2, 3]})
r1 = df.select("a").collect().to_pydict()
r2 = df.select("a").collect().to_pydict()
assert r1 == r2
\end{verbatim}}

\section{Extended Benchmark Details}

\begin{table*}[t]
\centering
\caption{UnseenSDK-Bench statistics per SDK. L0/L1/L2 denote curriculum levels. Property tests are generated automatically from type annotations.}
\label{tab:bench_stats}
\small
\begin{tabular}{lcccccc}
\toprule
\textbf{SDK} & \textbf{Category} & \textbf{Funcs} & \textbf{L0 Tasks} & \textbf{L1 Tasks} & \textbf{L2 Tasks} & \textbf{Prop.\ Tests} \\
\midrule
Daft & DataFrame & $156$ & $156$ & $312$ & $33$ & $624$ \\
LanceDB & Vector DB & $142$ & $142$ & $284$ & $33$ & $568$ \\
cuallee & Data Valid. & $87$ & $87$ & $174$ & $33$ & $348$ \\
Docling & Doc Proc. & $198$ & $198$ & $396$ & $33$ & $792$ \\
Marker & Doc Proc. & $112$ & $112$ & $224$ & $34$ & $448$ \\
Surya & Comp.\ Vision & $94$ & $94$ & $188$ & $33$ & $376$ \\
FastStream & Messaging & $234$ & $234$ & $468$ & $33$ & $936$ \\
Tracecat & Security & $312$ & $312$ & $624$ & $33$ & $1248$ \\
MCP Python & Protocol & $178$ & $178$ & $356$ & $33$ & $712$ \\
Vortex & Scientific & $145$ & $145$ & $290$ & $33$ & $580$ \\
libdebug & Debugging & $89$ & $89$ & $178$ & $33$ & $356$ \\
reflexio & Agent Frmwk & $124$ & $124$ & $248$ & $34$ & $496$ \\
\midrule
\textbf{Total} & -- & $1871$ & $1871$ & $3742$ & $398$ & $7484$ \\
\bottomrule
\end{tabular}
\end{table*}

\label{app:benchmark}

\subsection{Per-SDK Option Discovery Statistics}

\begin{table}[h]
\centering
\caption{Option discovery results per SDK. Composability density is $|E_{\mathrm{compose}}| / |F|^{2}$.}
\label{tab:option_discovery_stats}
\scriptsize
\setlength{\tabcolsep}{3pt}
\begin{tabular}{lccccc}
\toprule
\textbf{SDK} & \textbf{L0} & \textbf{L1 Cand.} & \textbf{L1 Filt.} & \textbf{Avg Len} & \textbf{Density} \\
\midrule
Daft & $156$ & $284$ & $73$ & $2.7$ & $0.041$ \\
LanceDB & $142$ & $241$ & $61$ & $2.5$ & $0.038$ \\
cuallee & $87$ & $159$ & $54$ & $2.4$ & $0.062$ \\
Docling & $198$ & $312$ & $48$ & $2.9$ & $0.029$ \\
Marker & $112$ & $187$ & $39$ & $2.6$ & $0.034$ \\
Surya & $94$ & $148$ & $35$ & $2.4$ & $0.036$ \\
FastStream & $234$ & $356$ & $52$ & $2.8$ & $0.024$ \\
Tracecat & $312$ & $402$ & $38$ & $2.5$ & $0.018$ \\
MCP Python & $178$ & $174$ & $24$ & $2.3$ & $0.011$ \\
Vortex & $145$ & $223$ & $47$ & $2.6$ & $0.030$ \\
libdebug & $89$ & $128$ & $29$ & $2.4$ & $0.028$ \\
reflexio & $124$ & $185$ & $33$ & $2.5$ & $0.026$ \\
\bottomrule
\end{tabular}
\end{table}

SDKs with higher composability density benefit more from the hierarchical approach: across the $12$ SDKs, composability density and the pass@1 improvement of Option-GRPO over Flat GRPO (Tables~\ref{tab:full_per_sdk_optgrpo} and~\ref{tab:full_per_sdk_flatgrpo}) have a Spearman correlation of $0.76$.

\subsection{Contamination Analysis}
\label{app:contamination}

To verify that the $12$ SDKs in UnseenSDK-Bench are functionally unseen by the Qwen3-32B pretraining corpus, we run two complementary checks. The first is a release-date comparison against the model's training cutoff (approximately February 2025, per the Qwen3 technical report~\cite{qwen3technicalreport}). The second is a memorisation probe: we query the frozen base model for SDK-specific identifiers and measure top-$1$ recall, comparing against two well-known control libraries (NumPy, PyTorch) whose presence in pretraining is uncontroversial.

\paragraph{Probe protocol.}
For each SDK, we construct $200$ identifier-completion prompts of the form \texttt{``In <SDK>, the <noun> that <verb> is called \_\_\_''}, where the noun and verb describe a specific class, method, or function in that SDK. We sample base-model completions at temperature $0$ and score the completion as correct if the top-$1$ token sequence matches the canonical identifier name (case-insensitive, ignoring leading dots or module prefixes). The $200$ prompts per SDK are split evenly across classes, methods, and free functions; identifiers are drawn uniformly from the SDK's public API. Control SDKs (NumPy, PyTorch) follow the same protocol with $200$ identifier prompts each. We run the same probe, unchanged ($200$ prompts per SDK), on GPT 5.5 and Claude Opus 4.8 and on the three earlier frontier models (GPT-4o, Claude 3.5 Sonnet, Gemini 2.0 Flash) through their APIs; per-SDK recall for the two current models is in Table~\ref{tab:contamination}, and the mean recall and unseen-SDK count of every model are in Table~\ref{tab:frontier}. Since these models are accessed through chat APIs, we wrap the same cloze prompts in a minimal instruction to reply with the identifier only and decode greedily; scoring is identical.

\begin{table}[h]
\centering
\caption{Contamination analysis. Release dates and the Cutoff column are relative to the Qwen3-32B training cutoff (approximately Feb 2025). Recall is top-$1$ identifier-completion accuracy on $200$ probes per SDK at temperature $0$, for the frozen Qwen3-32B and for the two current frontier models. $^{*}$Recall at or above the $25\%$ threshold: the SDK is functionally \emph{seen} by that model, so a comparison against that model on that SDK is a seen-SDK comparison rather than a like-for-like win. Controls are probed on the base model only.}
\label{tab:contamination}
\scriptsize
\setlength{\tabcolsep}{3pt}
\begin{tabular}{lccccc}
\toprule
 & & & \multicolumn{3}{c}{\textbf{Recall}} \\
\cmidrule(lr){4-6}
\textbf{SDK} & \textbf{Release} & \textbf{Cutoff} & \textbf{Qwen3} & \textbf{GPT 5.5} & \textbf{Opus 4.8} \\
\midrule
Daft        & 2022-08 & Pre  & $19.5\%$ & $48.5\%^{*}$ & $44.0\%^{*}$ \\
LanceDB     & 2023-03 & Pre  & $16.0\%$ & $59.0\%^{*}$ & $52.0\%^{*}$ \\
cuallee     & 2023-01 & Pre  & $12.5\%$ & $44.0\%^{*}$ & $41.5\%^{*}$ \\
Docling     & 2024-07 & Pre  & $\phantom{0}8.0\%$ & $46.0\%^{*}$ & $50.5\%^{*}$ \\
Marker      & 2024-02 & Pre  & $10.5\%$ & $52.0\%^{*}$ & $46.5\%^{*}$ \\
Surya       & 2024-01 & Pre  & $10.0\%$ & $21.5\%$ & $18.0\%$ \\
FastStream  & 2023-09 & Pre  & $17.5\%$ & $52.0\%^{*}$ & $47.0\%^{*}$ \\
Tracecat    & 2024-10 & Pre  & $\phantom{0}6.5\%$ & $22.5\%$ & $20.5\%$ \\
MCP Python  & 2024-11 & Pre  & $\phantom{0}4.0\%$ & $81.5\%^{*}$ & $84.0\%^{*}$ \\
Vortex      & 2024-06 & Pre  & $\phantom{0}7.0\%$ & $42.0\%^{*}$ & $52.5\%^{*}$ \\
libdebug    & 2024-03 & Pre  & $\phantom{0}8.5\%$ & $46.5\%^{*}$ & $41.0\%^{*}$ \\
reflexio    & 2024-10 & Pre  & $\phantom{0}5.5\%$ & $18.0\%$ & $34.5\%^{*}$ \\
\midrule
\textbf{UnseenSDK avg.} & --- & --- & $\mathbf{10.5\%}$ & $\mathbf{44.5\%}$ & $\mathbf{44.3\%}$ \\
\textbf{Unseen ($<25\%$)} & --- & --- & $12/12$ & $3/12$ & $2/12$ \\
\midrule
NumPy (ctrl) & 2006-01 & Pre & $91.5\%$ & --- & --- \\
PyTorch (ctrl) & 2016-09 & Pre & $88.0\%$ & --- & --- \\
\bottomrule
\end{tabular}
\end{table}

\paragraph{Interpretation.}
All $12$ SDKs are pre-cutoff under Qwen3's February~2025 corpus assembly date, so the unseen-SDK claim does not rest on post-cutoff novelty alone. Instead, it rests on the magnitude of recall. Per-SDK recall sits between $4.0\%$ and $19.5\%$, far below the NumPy ($91.5\%$) and PyTorch ($88.0\%$) controls. The highest-recall SDKs in our suite (Daft at $19.5\%$, FastStream at $17.5\%$) are the most established of the twelve, yet remain well below any reasonable threshold for ``the model knows this library.'' For comparison, prior work on novel-API code generation~\cite{hu2024exploracoder} uses a $25\%$ recall threshold to consider an API ``seen''; under that threshold, none of our $12$ SDKs qualifies.

The recall scores correlate with frozen-model pass@1 on UnseenSDK-Bench (Spearman $\rho = 0.70$, $p = 0.01$): the SDKs whose identifiers the base model partially recalls are also the ones the frozen model handles best. Even so, recall stays below $20\%$ on every SDK and frozen pass@1 averages $28.5\%$, so partial identifier memorisation does not amount to the compositional, type-correct API usage that the benchmark tasks require. We conclude that all $12$ SDKs in UnseenSDK-Bench are functionally unseen by Qwen3-32B, the model we train.

\paragraph{Per-model unseen-ness for current frontier models.}
Several of the SDKs, most notably MCP Python but also Daft, LanceDB, and Docling, have grown substantially in visibility since early 2025 and are likely represented in current frontier training corpora. Under the $25\%$ threshold, $9$ of the $12$ SDKs are functionally seen by GPT 5.5 and $10$ of $12$ by Claude Opus 4.8 (Table~\ref{tab:contamination}), so a rerun on the full suite would mostly measure seen-SDK performance rather than the unseen-SDK setting this paper studies. Seen-ness is also model-specific: reflexio is below the threshold for GPT 5.5 but above it for Claude Opus 4.8, which is why recall is reported per model. Two SDKs, Surya and Tracecat, remain below the threshold for both models, and the like-for-like comparison with current frontier models is restricted to these two (Table~\ref{tab:frontier_unseen}).

\section{Extended Experimental Results}
\label{app:results}

\subsection{Full Per-SDK Results}
\label{app:full_per_sdk}

Table~\ref{tab:full_per_sdk_optgrpo} reports all six evaluation metrics for Option-GRPO on each of the 12 SDKs. The option reuse rate (OR) is the fraction of SDK API calls in passing solutions that fall inside a certified Level-1 option pattern. Table~\ref{tab:full_per_sdk_flatgrpo} reports the same for Flat GRPO, the non-hierarchical critic-free RL baseline that uses our layered execution reward.

\begin{table*}[h]
\centering
\caption{Full per-SDK results for Option-GRPO: pass@1, pass@10, code length CL, API efficiency AE (lower is better), idiomaticity ID (higher is better), option reuse rate OR (higher is better).}
\label{tab:full_per_sdk_optgrpo}
\footnotesize
\setlength{\tabcolsep}{4pt}
\begin{tabular}{lccccccc}
\toprule
\textbf{SDK} & \textbf{Category} & \textbf{p@1} & \textbf{p@10} & \textbf{CL} $\downarrow$ & \textbf{AE} $\downarrow$ & \textbf{ID} $\uparrow$ & \textbf{OR} $\uparrow$ \\
\midrule
cuallee     & Data Valid.  & $78.2$ & $91.0$ & $39.4$ & $2.3$ & $0.89$ & $0.80$ \\
LanceDB     & Vector DB    & $74.6$ & $87.4$ & $42.7$ & $2.7$ & $0.83$ & $0.70$ \\
Marker      & Doc Proc.    & $74.1$ & $86.8$ & $44.5$ & $3.0$ & $0.82$ & $0.66$ \\
Surya       & Comp.~Vision & $63.2$ & $78.4$ & $51.9$ & $3.6$ & $0.80$ & $0.62$ \\
Daft        & DataFrame    & $72.9$ & $85.0$ & $45.8$ & $2.7$ & $0.83$ & $0.72$ \\
Vortex      & Scientific   & $60.6$ & $77.0$ & $49.6$ & $3.1$ & $0.70$ & $0.63$ \\
libdebug    & Debugging    & $60.2$ & $76.3$ & $54.6$ & $3.1$ & $0.76$ & $0.53$ \\
FastStream  & Messaging    & $57.1$ & $74.1$ & $55.4$ & $3.5$ & $0.71$ & $0.51$ \\
Docling     & Doc Proc.    & $54.5$ & $71.3$ & $61.0$ & $3.4$ & $0.67$ & $0.43$ \\
reflexio    & Agent Frmwk  & $51.4$ & $69.9$ & $57.6$ & $3.7$ & $0.71$ & $0.42$ \\
Tracecat    & Security     & $46.8$ & $63.1$ & $58.9$ & $3.8$ & $0.58$ & $0.39$ \\
MCP Python  & Protocol     & $43.2$ & $61.3$ & $62.6$ & $3.5$ & $0.58$ & $0.31$ \\
\midrule
\textbf{Average} & --      & $\mathbf{61.4}$ & $\mathbf{76.8}$ & $\mathbf{52.0}$ & $\mathbf{3.2}$ & $\mathbf{0.74}$ & $\mathbf{0.56}$ \\
\bottomrule
\end{tabular}
\end{table*}

\begin{table*}[h]
\centering
\caption{Full per-SDK results for Flat GRPO (no hierarchy, layered execution reward). Same metrics as Table~\ref{tab:full_per_sdk_optgrpo}.}
\label{tab:full_per_sdk_flatgrpo}
\footnotesize
\setlength{\tabcolsep}{4pt}
\begin{tabular}{lccccccc}
\toprule
\textbf{SDK} & \textbf{Category} & \textbf{p@1} & \textbf{p@10} & \textbf{CL} $\downarrow$ & \textbf{AE} $\downarrow$ & \textbf{ID} $\uparrow$ & \textbf{OR} $\uparrow$ \\
\midrule
cuallee     & Data Valid.  & $67.2$ & $83.0$ & $61.6$ & $4.2$ & $0.52$ & $0.38$ \\
LanceDB     & Vector DB    & $63.8$ & $79.7$ & $64.7$ & $4.5$ & $0.45$ & $0.31$ \\
Marker      & Doc Proc.    & $64.8$ & $81.0$ & $62.4$ & $4.5$ & $0.49$ & $0.32$ \\
Surya       & Comp.~Vision & $55.8$ & $70.9$ & $70.2$ & $5.0$ & $0.44$ & $0.25$ \\
Daft        & DataFrame    & $59.9$ & $76.3$ & $66.2$ & $4.6$ & $0.49$ & $0.35$ \\
Vortex      & Scientific   & $51.7$ & $67.6$ & $71.0$ & $4.8$ & $0.38$ & $0.25$ \\
libdebug    & Debugging    & $51.2$ & $66.8$ & $73.3$ & $5.0$ & $0.41$ & $0.26$ \\
FastStream  & Messaging    & $49.3$ & $65.0$ & $73.4$ & $5.0$ & $0.38$ & $0.20$ \\
Docling     & Doc Proc.    & $45.2$ & $60.9$ & $80.0$ & $5.1$ & $0.35$ & $0.21$ \\
reflexio    & Agent Frmwk  & $42.8$ & $58.1$ & $77.3$ & $5.3$ & $0.37$ & $0.17$ \\
Tracecat    & Security     & $38.2$ & $54.6$ & $78.4$ & $5.5$ & $0.30$ & $0.15$ \\
MCP Python  & Protocol     & $42.5$ & $56.9$ & $77.1$ & $5.3$ & $0.34$ & $0.15$ \\
\midrule
\textbf{Average} & --      & $\mathbf{52.7}$ & $\mathbf{68.4}$ & $\mathbf{71.3}$ & $\mathbf{4.9}$ & $\mathbf{0.41}$ & $\mathbf{0.25}$ \\
\bottomrule
\end{tabular}
\end{table*}

Aggregate deltas: pass@1 improvement of Option-GRPO over Flat GRPO ranges from $+0.7$ points on MCP Python to $+13.0$ points on Daft, with a mean improvement of $+8.7$ points. CL drops by an average of $27\%$, AE by $34.7\%$, ID rises by $0.33$, and OR rises by $0.31$. The largest improvements are on data-processing SDKs (cuallee, LanceDB, Daft) where compositional structure in the parameter-flow graph is densest; the smallest improvement is on MCP Python, where type-based composability is weakest, and the two lowest-density SDKs (MCP Python, Tracecat) have the lowest absolute pass@1, matching the option-discovery density statistics in Table~\ref{tab:option_discovery_stats}.

\subsection{Convergence Curves}

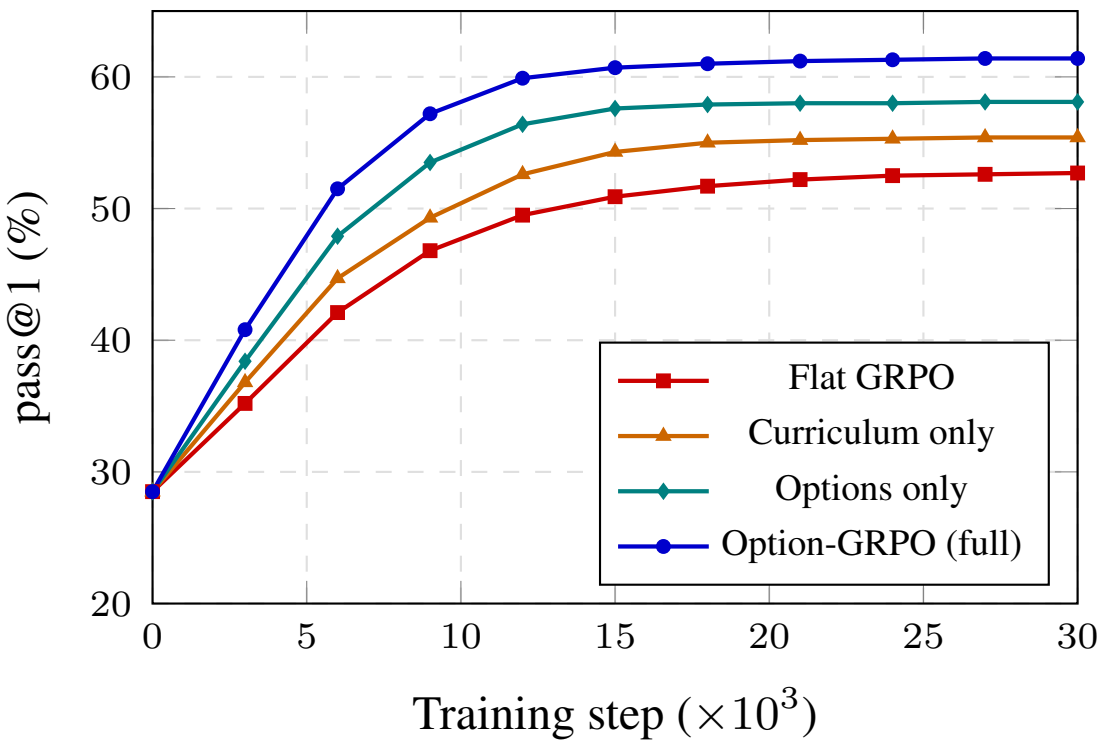
\begin{figure}[h]
\centering
\begin{tikzpicture}
\begin{axis}[
    width=7.7cm, height=5.5cm,
    xlabel={Training step ($\times 10^3$)},
    ylabel={pass@1 (\%)},
    xmin=0, xmax=30,
    ymin=20, ymax=65,
    xtick={0,5,10,15,20,25,30},
    ytick={20,30,40,50,60},
    grid=major, grid style={dashed,gray!25},
    legend pos=south east, legend style={font=\scriptsize},
    tick label style={font=\scriptsize},
    label style={font=\small},
]
\addplot[thick, color=red!80!black, mark=square*, mark size=1pt] coordinates {
(0, 28.5) (3, 35.2) (6, 42.1) (9, 46.8) (12, 49.5) (15, 50.9) (18, 51.7) (21, 52.2) (24, 52.5) (27, 52.6) (30, 52.7)};
\addlegendentry{Flat GRPO}
\addplot[thick, color=orange!80!black, mark=triangle*, mark size=1pt] coordinates {
(0, 28.5) (3, 36.8) (6, 44.7) (9, 49.3) (12, 52.6) (15, 54.3) (18, 55.0) (21, 55.2) (24, 55.3) (27, 55.4) (30, 55.4)};
\addlegendentry{Curriculum only}
\addplot[thick, color=teal, mark=diamond*, mark size=1pt] coordinates {
(0, 28.5) (3, 38.4) (6, 47.9) (9, 53.5) (12, 56.4) (15, 57.6) (18, 57.9) (21, 58.0) (24, 58.0) (27, 58.1) (30, 58.1)};
\addlegendentry{Options only}
\addplot[thick, color=blue!80!black, mark=*, mark size=1pt] coordinates {
(0, 28.5) (3, 40.8) (6, 51.5) (9, 57.2) (12, 59.9) (15, 60.7) (18, 61.0) (21, 61.2) (24, 61.3) (27, 61.4) (30, 61.4)};
\addlegendentry{Option-GRPO (full)}
\end{axis}
\end{tikzpicture}
\caption{Convergence curves averaged across 12 SDKs. Option-GRPO reaches the flat-GRPO terminal pass@1 in roughly one-third the steps and continues to improve.}
\label{fig:convergence}
\end{figure}

Figure~\ref{fig:convergence} shows pass@1 against training step for the four configurations of the hierarchy ablation (Table~\ref{tab:hierarchy_ablation}), averaged across the 12 SDKs. Option-GRPO reaches the flat-GRPO terminal pass@1 ($52.7\%$) before step $9{,}000$ (under one-third of the $30{,}000$-step budget) and continues climbing to $61.4\%$ by step $30{,}000$. Curriculum-only and options-only configurations lie between the extremes, with options-only slightly ahead at most checkpoints, consistent with the larger marginal contribution of option-level action structure.

\subsection{Option Hierarchy Visualisations}
\label{app:hierarchy_viz}

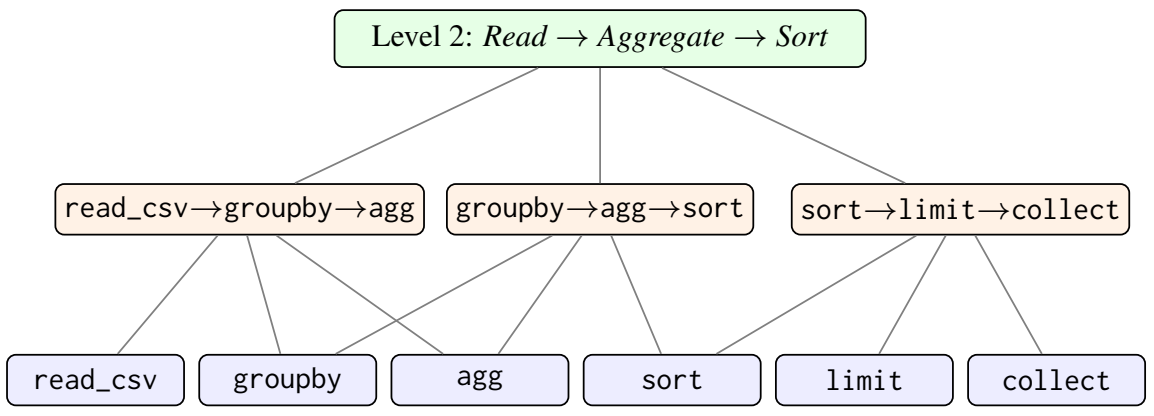
\begin{figure}[h]
\centering
\resizebox{\columnwidth}{!}{
\begin{tikzpicture}[
    every node/.style={font=\scriptsize},
    l0/.style={draw, rounded corners=2pt, fill=blue!7, inner sep=2pt, minimum width=14mm, minimum height=4mm},
    l1/.style={draw, rounded corners=2pt, fill=orange!10, inner sep=2pt, minimum width=24mm, minimum height=4mm},
    l2/.style={draw, rounded corners=2pt, fill=green!10, inner sep=2pt, minimum width=42mm, minimum height=4.5mm},
    arr/.style={-, gray, thin},
    yscale=0.9, xscale=0.95
]
\node[l2] (root) at (0, 3.5) {Level 2: \emph{Read $\to$ Aggregate $\to$ Sort}};
\node[l1] (l1a) at (-3, 2) {\texttt{read\_csv$\to$groupby$\to$agg}};
\node[l1] (l1b) at ( 0, 2) {\texttt{groupby$\to$agg$\to$sort}};
\node[l1] (l1c) at ( 3, 2) {\texttt{sort$\to$limit$\to$collect}};
\node[l0] (a) at (-4.2, 0.5) {\texttt{read\_csv}};
\node[l0] (b) at (-2.6, 0.5) {\texttt{groupby}};
\node[l0] (c) at (-1.0, 0.5) {\texttt{agg}};
\node[l0] (d) at ( 0.6, 0.5) {\texttt{sort}};
\node[l0] (e) at ( 2.2, 0.5) {\texttt{limit}};
\node[l0] (f) at ( 3.8, 0.5) {\texttt{collect}};
\draw[arr] (root) -- (l1a);
\draw[arr] (root) -- (l1b);
\draw[arr] (root) -- (l1c);
\draw[arr] (l1a) -- (a);
\draw[arr] (l1a) -- (b);
\draw[arr] (l1a) -- (c);
\draw[arr] (l1b) -- (b);
\draw[arr] (l1b) -- (c);
\draw[arr] (l1b) -- (d);
\draw[arr] (l1c) -- (d);
\draw[arr] (l1c) -- (e);
\draw[arr] (l1c) -- (f);
\end{tikzpicture}
}
\caption{Discovered option hierarchy for Daft (excerpt). Blue: Level-0 atomic options (one per SDK function). Orange: Level-1 compositional options (centroid paths of length 3 from Algorithm~\ref{alg:option_discovery}). Green: Level-2 program-level composition learned at Phase~3. Edges denote type-composability inclusion. The full Daft hierarchy contains $156$ Level-0 options and $73$ Level-1 options; we show only the chain underlying a representative Level-2 program.}
\label{fig:option_tree_daft}
\end{figure}

Figure~\ref{fig:option_tree_daft} shows the discovered option hierarchy for Daft as a layered graph. Level-0 options (atomic SDK calls) form the leaves; Level-1 options (compositional patterns mined from $G_{\mathrm{flow}}$) form the internal nodes; Level-2 patterns (full programs assembled during Phase~3) sit at the root. Edges denote type composability ($o_i \prec o_j$ iff $\mathcal{T}_i^{\mathrm{out}} \cap \mathcal{T}_j^{\mathrm{in}} \neq \emptyset$). Three observations: (i)~the Level-1 layer is dominated by the canonical \emph{load $\to$ transform $\to$ collect} pattern, which appears as the highest-weight cluster in Algorithm~\ref{alg:option_discovery}; (ii)~the parameter-flow graph is densely connected within each domain cluster (DataFrame operations chain freely among themselves) but sparse across clusters; (iii)~options containing \texttt{collect} act as natural terminators because their output type is a Python list, not a Daft expression.

Analogous hierarchies for LanceDB and cuallee follow the same structural pattern, with vector-search chains (\texttt{connect$\to$create\_table$\to$add$\to$search}) dominating the LanceDB Level-1 layer and validation-rule chains (\texttt{Check$\to$is\_complete$\to$is\_unique$\to$validate}) dominating cuallee's. Across the three data-processing SDKs, the discovered Level-1 options correspond exactly to the canonical usage idioms in each library's official tutorial, confirming that the parameter-flow graph captures the intended compositional structure rather than spurious type coincidences.

\subsection{Code Quality Examples: Slop Reduction Side-by-Side}
\label{app:slop_examples}

This subsection presents two representative side-by-side comparisons of Flat GRPO output and Option-GRPO output on the same task. Annotations report token count $|c|$, number of SDK API calls $N_{\mathrm{api}}$, number of redundant intermediate variables $N_{\mathrm{redundant}}$, and option description length $\mathrm{ODL}_{\mathcal{H}}(c)$. The examples span easy (Daft) and medium (LanceDB) difficulty.

\paragraph{Example 1 (easy, Daft).} Task: \textit{``Read \texttt{employees.csv} and return rows where \texttt{salary} exceeds 75{,}000.''}

\noindent\textbf{Flat GRPO} ($|c|{=}82$, $N_{\mathrm{api}}{=}5$, $N_{\mathrm{redundant}}{=}3$, $\mathrm{ODL}{=}\infty$ since the decomposition uses non-Daft calls):
{\footnotesize\begin{verbatim}
import daft
import pandas as pd     # unused
path = "employees.csv"
df_raw = daft.read_csv(path)
df = df_raw.collect()
result_list = []
for row in df.iter_rows():
    if row["salary"] > 75000:
        result_list.append(row)
result_df = daft.from_pylist(result_list)
final = result_df.collect()
\end{verbatim}}

\noindent\textbf{Option-GRPO} ($|c|{=}24$, $N_{\mathrm{api}}{=}4$, $N_{\mathrm{redundant}}{=}0$, $\mathrm{ODL}{=}1$, one Level-1 option \texttt{Read$\to$Filter$\to$Collect}):
{\footnotesize\begin{verbatim}
import daft
df = daft.read_csv("employees.csv")
result = (df
    .where(daft.col("salary") > 75000)
    .collect())
\end{verbatim}}

\noindent The Flat-GRPO solution introduces an unused \texttt{pandas} import, materializes a list intermediate, and reconstructs the DataFrame; the Option-GRPO solution invokes a single discovered option that chains \texttt{read\_csv}, \texttt{where}, \texttt{collect}.

\paragraph{Example 2 (medium, LanceDB).} Task: \textit{``Create a LanceDB table from a list of vectors and return the 5 nearest neighbours to a query vector.''}

\noindent\textbf{Flat GRPO} ($|c|{=}146$, $N_{\mathrm{api}}{=}7$: five LanceDB calls plus the \texttt{pandas} and \texttt{numpy} constructors, $N_{\mathrm{redundant}}{=}4$, $\mathrm{ODL}{=}\infty$):
{\footnotesize\begin{verbatim}
import lancedb
import numpy as np
import pandas as pd
data = [
  {"vec":[0.1,0.2,0.3],"id":1},
  {"vec":[0.4,0.5,0.6],"id":2}]
db = lancedb.connect("/tmp/db")
df_data = pd.DataFrame(data)
table_obj = db.create_table(
    "vectors", data=df_data)
query_arr = np.array([0.1,0.2,0.3])
search_obj = table_obj.search(
    query_arr.tolist())
results_raw = (search_obj
    .limit(5).to_pandas())
results = results_raw.to_dict("records")
\end{verbatim}}

\noindent\textbf{Option-GRPO} ($|c|{=}48$, $N_{\mathrm{api}}{=}5$, $N_{\mathrm{redundant}}{=}0$, $\mathrm{ODL}{=}2$; two Level-1 options \texttt{Connect$\to$CreateTable} and \texttt{Search$\to$Limit$\to$Materialize}):
{\footnotesize\begin{verbatim}
import lancedb
db = lancedb.connect("/tmp/db")
table = db.create_table("vectors", data=[
    {"vec":[0.1,0.2,0.3],"id":1},
    {"vec":[0.4,0.5,0.6],"id":2}])
results = (table
    .search([0.1,0.2,0.3])
    .limit(5)
    .to_list())
\end{verbatim}}

\noindent Flat GRPO routes the data through \texttt{pandas} and \texttt{numpy} unnecessarily, introducing four redundant intermediate bindings (\texttt{df\_data}, \texttt{query\_arr}, \texttt{search\_obj}, \texttt{results\_raw}); Option-GRPO uses the native LanceDB list and array APIs.

\paragraph{Aggregate slop reduction across the two examples.} Token counts drop $71\%$ (82$\to$24) and $67\%$ (146$\to$48) respectively. Non-SDK calls drop from 5 to 0 across the two examples. Redundant intermediates drop from 7 total to 0 total. ODL drops from $\infty$ (no valid decomposition under the option set) to small finite values (1--2 options). These two examples are representative; Spearman correlations across all $5{,}042$ passing solutions are reported in Figure~\ref{fig:odl_quality}.

\subsection{Gradient Variance Analysis}

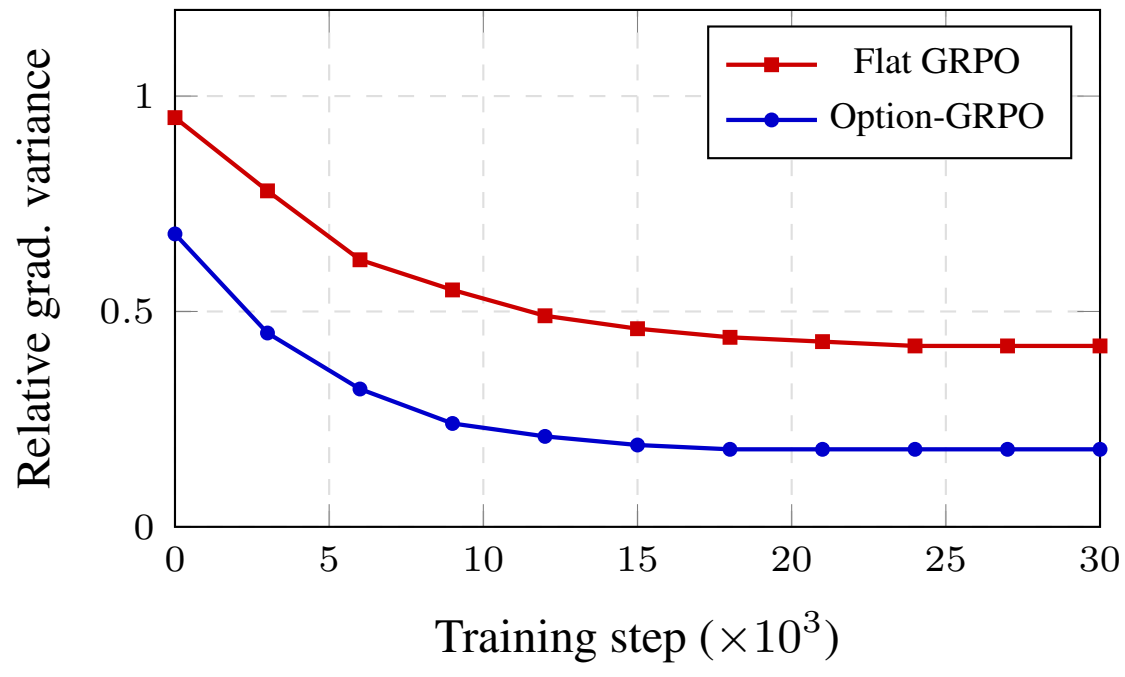
\begin{figure}[h]
\centering
\begin{tikzpicture}
\begin{axis}[
    width=7.7cm, height=5cm,
    xlabel={Training step ($\times 10^3$)},
    ylabel={Relative grad. variance},
    xmin=0, xmax=30,
    ymin=0, ymax=1.2,
    xtick={0,5,10,15,20,25,30},
    grid=major, grid style={dashed,gray!25},
    legend pos=north east, legend style={font=\scriptsize},
    tick label style={font=\scriptsize},
    label style={font=\small},
]
\addplot[thick, color=red!80!black, mark=square*, mark size=1pt] coordinates {
(0, 0.95) (3, 0.78) (6, 0.62) (9, 0.55) (12, 0.49) (15, 0.46) (18, 0.44) (21, 0.43) (24, 0.42) (27, 0.42) (30, 0.42)};
\addlegendentry{Flat GRPO}
\addplot[thick, color=blue!80!black, mark=*, mark size=1pt] coordinates {
(0, 0.68) (3, 0.45) (6, 0.32) (9, 0.24) (12, 0.21) (15, 0.19) (18, 0.18) (21, 0.18) (24, 0.18) (27, 0.18) (30, 0.18)};
\addlegendentry{Option-GRPO}
\end{axis}
\end{tikzpicture}
\caption{Per-step relative gradient norm variance (averaged over 12 SDKs). Option-GRPO shows roughly $2.3\times$ lower variance at the end of training (about $1.5\times$ in SNR), in the direction predicted by Proposition~\ref{prop:credit} but smaller than its $\sqrt{T/K}$ scaling.}
\label{fig:grad_variance}
\end{figure}

Figure~\ref{fig:grad_variance} plots the per-step relative gradient norm variance (computed as the ratio of the gradient-norm variance to the squared mean gradient norm, equivalent to the inverse SNR squared) for flat GRPO and Option-GRPO. Option-GRPO maintains lower variance throughout training, with the gap widening from $1.4\times$ at initialization to about $2.4\times$ by step $18{,}000$ and $2.3\times$ at the end of training. The reduction is in the direction predicted by Proposition~\ref{prop:credit} but smaller than its scaling: a $2.3\times$ lower relative variance corresponds to an SNR gain of about $1.5\times$, against $\sqrt{T/K} \approx 4.5\times$ for $T{=}100$, $K{=}5$. The assumptions of the proposition (reward-correlated signal confined to option-decision positions, conditionally independent positions) are idealizations, so it should be read as giving the direction and scaling of the effect, not its measured size.

\subsection{Full Ablation and Per-SDK Tables}
\label{app:rl_methods}
\label{app:frontier}

\begin{table}[h]
\centering
\caption{Hierarchy ablation. Both options and curriculum contribute; their combination yields the largest gains. Rewards differ across rows: Flat GRPO uses the layered execution reward only; Curric.\ only and Full use the same schedule (execution reward, plus the quality term from Phase~2 and the consistency term in Phase~3); Options only uses the full reward throughout.}
\label{tab:hierarchy_ablation}
\footnotesize
\setlength{\tabcolsep}{3pt}
\begin{tabular}{lcccc}
\toprule
\textbf{Config.} & \textbf{p@1} & \textbf{CL} $\downarrow$ & \textbf{ID} $\uparrow$ & \textbf{Conv. (steps)} \\
\midrule
Flat GRPO & $52.7$ & $71.3$ & $0.41$ & $24{,}800$ \\
Curric.\ only & $55.4$ & $66.8$ & $0.49$ & $19{,}200$ \\
Options only & $58.1$ & $58.7$ & $0.66$ & $15{,}600$ \\
\textbf{Full (Option-GRPO)} & $\mathbf{61.4}$ & $\mathbf{52.0}$ & $\mathbf{0.74}$ & $\mathbf{12{,}400}$ \\
\bottomrule
\end{tabular}
\end{table}

\begin{table}[h]
\centering
\caption{Further ablations (pass@1, \%, averaged over the $12$ SDKs). Top: curriculum levels used in training. Bottom: reward function.}
\label{tab:further_ablations}
\footnotesize
\setlength{\tabcolsep}{4pt}
\begin{tabular}{lc}
\toprule
\textbf{Config.} & \textbf{p@1} \\
\midrule
\multicolumn{2}{l}{\emph{Option level}} \\
Level~0 only & $53.9$ \\
Levels~0 + 1 & $58.6$ \\
\textbf{Full (Levels~0 + 1 + 2)} & $\mathbf{61.4}$ \\
\midrule
\multicolumn{2}{l}{\emph{Reward}} \\
Binary reward only & $34.2$ \\
\textbf{Layered + quality + consistency (full)} & $\mathbf{61.4}$ \\
\bottomrule
\end{tabular}
\end{table}

\paragraph{Further ablations.} Table~\ref{tab:further_ablations} reports the option-level and reward-function ablations. For credit assignment, combining the sequence-level and option-level advantages outperforms either alone by $4.1$--$4.6$ pass@1 points. In the curriculum-pacing sweep, the adaptive setting $\rho_{\mathrm{adv}}{=}0.5$ is best. In the group-size sweep, $G{=}16$ is the Pareto knee. Per-SDK results are in Table~\ref{tab:per_sdk}.

\begin{table}[h]
\centering
\caption{Cross-SDK option transfer (pass@1 \%). Rows: source SDK; columns: target SDK.}
\label{tab:transfer}
\small
\setlength{\tabcolsep}{4pt}
\begin{tabular}{lccc}
\toprule
\textbf{Source} $\to$ \textbf{Target} & \textbf{Daft} & \textbf{LanceDB} & \textbf{cuallee} \\
\midrule
No transfer (from scratch) & $72.9$ & $74.6$ & $78.2$ \\
Daft & -- & $76.2$ & $78.9$ \\
LanceDB & $74.4$ & -- & $79.5$ \\
cuallee & $73.6$ & $75.6$ & -- \\
\bottomrule
\end{tabular}
\end{table}

\begin{table}[h]
\centering
\caption{Comparison with prior RL methods on UnseenSDK-Bench. $^{*}$Adapted to use our layered execution reward in place of unit tests. All use the same Qwen3-32B base.}
\label{tab:rl_comparison}
\footnotesize
\setlength{\tabcolsep}{3pt}
\begin{tabular}{lccccc}
\toprule
\textbf{Method} & \textbf{p@1} & \textbf{p@10} & \textbf{Critic} & \textbf{CL} $\downarrow$ & \textbf{ID} $\uparrow$ \\
\midrule
CodeRL-adapted$^{*}$ & $46.8$ & $63.2$ & Yes & $79.5$ & $0.38$ \\
PPOCoder-adapted$^{*}$ & $48.5$ & $64.7$ & Yes & $76.1$ & $0.40$ \\
StepCoder-adapted$^{*}$ & $51.3$ & $67.9$ & Yes & $72.8$ & $0.45$ \\
ExploraCoder & $50.2$ & $67.5$ & No & $78.4$ & $0.43$ \\
ReCode & $55.1$ & $71.2$ & No & $74.1$ & $0.51$ \\
Flat GRPO & $52.7$ & $68.4$ & No & $71.3$ & $0.41$ \\
\textbf{Option-GRPO} & $\mathbf{61.4}$ & $\mathbf{76.8}$ & \textbf{No} & $\mathbf{52.0}$ & $\mathbf{0.74}$ \\
\bottomrule
\end{tabular}
\end{table}

\begin{table}[h]
\centering
\caption{Comparison with frontier LLMs on the full $12$-SDK suite. The frontier LLMs receive the same top-10 RAG context (source-code chunks). Option-GRPO is shown without retrieval, which makes the comparison conservative: adding RAG improves Option-GRPO to $63.7$ pass@1 and $78.5$ pass@10 (Table~\ref{tab:complementarity}). The other Qwen3-32B rows repeat Table~\ref{tab:main}. Recall is mean top-$1$ identifier recall over the $12$ SDKs under the probe of Appendix~\ref{app:contamination} (for ReCode and Option-GRPO, that of the Qwen3-32B base); Unseen is the number of SDKs below the $25\%$ threshold for that model. A full-suite row is a like-for-like comparison only on the SDKs unseen to that model. GPT 5.5 and Claude Opus 4.8 are functionally seen on most of the suite and are compared only on the SDKs unseen to both (Table~\ref{tab:frontier_unseen}).}
\label{tab:frontier}
\scriptsize
\setlength{\tabcolsep}{3pt}
\begin{tabular}{lccccc}
\toprule
\textbf{Method} & \textbf{Recall} & \textbf{Unseen} & \textbf{p@1} & \textbf{p@10} & $\Delta$ \\
\midrule
Frozen Qwen3-32B & $10.5\%$ & $12/12$ & $28.5$ & $42.1$ & $-32.9$ \\
GPT-4o & $27\%$ & $8/12$ & $54.6$ & $69.4$ & $-6.8$ \\
Claude 3.5 Sonnet & $35\%$ & $6/12$ & $57.2$ & $71.8$ & $-4.2$ \\
Gemini 2.0 Flash & $32\%$ & $7/12$ & $51.8$ & $67.1$ & $-9.6$ \\
GPT 5.5 & $44.5\%$ & $3/12$ & \multicolumn{3}{c}{Table~\ref{tab:frontier_unseen}} \\
Claude Opus 4.8 & $44.3\%$ & $2/12$ & \multicolumn{3}{c}{Table~\ref{tab:frontier_unseen}} \\
ReCode (32B) & $10.5\%$ & $12/12$ & $55.1$ & $71.2$ & $-6.3$ \\
\textbf{Option-GRPO (32B)} & $10.5\%$ & $12/12$ & $\mathbf{61.4}$ & $\mathbf{76.8}$ & -- \\
\bottomrule
\end{tabular}
\end{table}

\begin{table}[h]
\centering
\caption{Like-for-like comparison with current frontier models on the two SDKs that are functionally unseen to both (recall below $25\%$, Table~\ref{tab:contamination}). Same protocol as Table~\ref{tab:frontier}: the frontier models receive the top-10 source-chunk RAG context and Option-GRPO is evaluated without retrieval; identical prompts, $n{=}10$ samples, temperature $0.8$. Ref.\ p@1 is pass@1 under the stricter reference-output-equivalence scorer.}
\label{tab:frontier_unseen}
\footnotesize
\setlength{\tabcolsep}{3pt}
\begin{tabular}{llccc}
\toprule
\textbf{SDK} & \textbf{Metric} & \textbf{GPT 5.5} & \textbf{Opus 4.8} & \textbf{Opt-GRPO} \\
\midrule
Surya    & pass@1    & $55.3$ & $57.5$ & $\mathbf{63.2}$ \\
Surya    & pass@10   & $71.8$ & $73.7$ & $\mathbf{78.4}$ \\
Surya    & Ref.\ p@1 & $45.3$ & $44.7$ & $\mathbf{56.2}$ \\
\midrule
Tracecat & pass@1    & $38.4$ & $42.1$ & $\mathbf{46.8}$ \\
Tracecat & pass@10   & $45.8$ & $50.4$ & $\mathbf{63.1}$ \\
Tracecat & Ref.\ p@1 & $31.8$ & $30.2$ & $\mathbf{37.8}$ \\
\bottomrule
\end{tabular}
\end{table}

Option-GRPO leads in every cell of Table~\ref{tab:frontier_unseen}. The pass@1 margin over the stronger current model is $+5.7$ on Surya and $+4.7$ on Tracecat ($95\%$ CI $+3.6$ to $+8.7$ and $+3.9$ to $+7.3$, paired bootstrap over tasks), and the pass@10 margin on Tracecat is $+12.7$. Option-GRPO also leads under the stricter reference-output-equivalence scorer.

\begin{table*}[h]
\centering
\caption{Per-SDK results on UnseenSDK-Bench (pass@1 \%). ExpC.~= ExploraCoder.}
\label{tab:per_sdk}
\small
\begin{tabular}{lccccccc}
\toprule
\textbf{SDK} & \textbf{Frozen} & \textbf{ExpC.} & \textbf{ReCode} & \textbf{Flat GRPO} & \textbf{Opt-GRPO} & $\Delta$ \textbf{vs ReCode} & \textbf{Category} \\
\midrule
cuallee & $47.1$ & $65.2$ & $69.6$ & $67.2$ & $\mathbf{78.2}$ & $+8.6$ & Data Valid. \\
LanceDB & $40.2$ & $62.0$ & $66.7$ & $63.8$ & $\mathbf{74.6}$ & $+7.9$ & Vector DB \\
Marker & $43.9$ & $63.1$ & $67.5$ & $64.8$ & $\mathbf{74.1}$ & $+6.6$ & Doc Proc. \\
Surya & $34.8$ & $54.8$ & $58.2$ & $55.8$ & $\mathbf{63.2}$ & $+5.0$ & Comp.\ Vision \\
Daft & $34.2$ & $58.1$ & $64.2$ & $59.9$ & $\mathbf{72.9}$ & $+8.7$ & DataFrame \\
Vortex & $28.0$ & $50.1$ & $54.0$ & $51.7$ & $\mathbf{60.6}$ & $+6.6$ & Scientific \\
libdebug & $27.4$ & $48.8$ & $53.6$ & $51.2$ & $\mathbf{60.2}$ & $+6.6$ & Debugging \\
FastStream & $24.7$ & $46.8$ & $50.8$ & $49.3$ & $\mathbf{57.1}$ & $+6.3$ & Messaging \\
Docling & $20.8$ & $43.3$ & $48.8$ & $45.2$ & $\mathbf{54.5}$ & $+5.7$ & Doc Proc. \\
reflexio & $18.4$ & $40.9$ & $45.6$ & $42.8$ & $\mathbf{51.4}$ & $+5.8$ & Agent Frmwk \\
Tracecat & $11.6$ & $34.0$ & $39.8$ & $38.2$ & $\mathbf{46.8}$ & $+7.0$ & Security \\
MCP Python & $10.9$ & $35.3$ & $42.4$ & $42.5$ & $\mathbf{43.2}$ & $+0.8$ & Protocol \\
\midrule
\textbf{Average} & $28.5$ & $50.2$ & $55.1$ & $52.7$ & $\mathbf{61.4}$ & $+6.3$ & -- \\
\bottomrule
\end{tabular}
\end{table*}

\begin{table}[h]
\centering
\caption{Complementarity with retrieval at inference time.}
\label{tab:complementarity}
\small
\begin{tabular}{lcc}
\toprule
\textbf{Configuration} & \textbf{p@1} & \textbf{p@10} \\
\midrule
RAG only (32B) & $41.3$ & $56.8$ \\
DocPrompting only (32B) & $44.8$ & $60.2$ \\
Option-GRPO (no retrieval) & $61.4$ & $76.8$ \\
Option-GRPO + RAG & $\mathbf{63.7}$ & $\mathbf{78.5}$ \\
Option-GRPO + DocPrompting & $63.2$ & $78.1$ \\
\bottomrule
\end{tabular}
\end{table}

\label{sec:complementarity}

\subsection{Empirical Validation of the Compression Results}
\label{app:compression_validation}

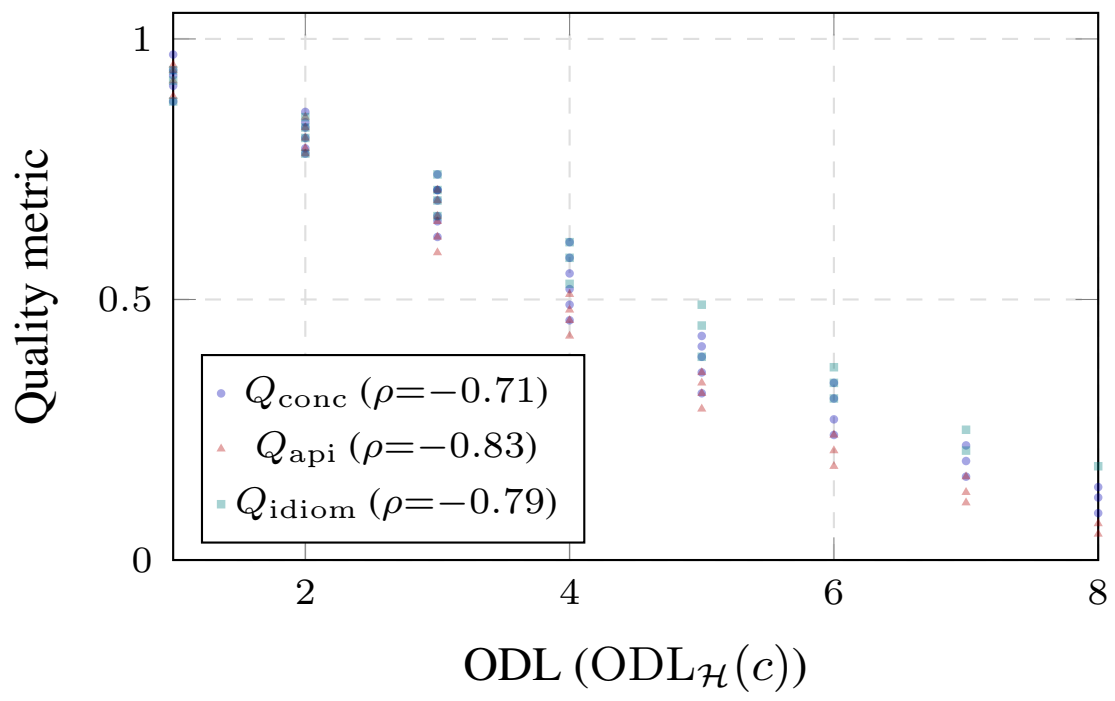
\begin{figure}[h]
\centering
\begin{tikzpicture}
\begin{axis}[
    width=7.7cm, height=5.2cm,
    xlabel={ODL ($\mathrm{ODL}_{\mathcal{H}}(c)$)},
    ylabel={Quality metric},
    xmin=1, xmax=8,
    ymin=0, ymax=1.05,
    grid=major, grid style={dashed,gray!25},
    legend pos=south west, legend style={font=\scriptsize},
    tick label style={font=\scriptsize},
    label style={font=\small},
]
\addplot[only marks, mark=*, mark size=0.6pt, color=blue!70!black, opacity=0.35] coordinates {
(1, 0.97) (1, 0.91) (1, 0.88) (1, 0.94) (1, 0.93) (2, 0.86) (2, 0.79) (2, 0.83) (2, 0.78) (2, 0.81) (2, 0.84) (3, 0.74) (3, 0.66) (3, 0.71) (3, 0.69) (3, 0.62) (3, 0.65) (3, 0.71) (4, 0.55) (4, 0.49) (4, 0.61) (4, 0.52) (4, 0.46) (4, 0.58) (5, 0.41) (5, 0.36) (5, 0.43) (5, 0.32) (5, 0.39) (6, 0.27) (6, 0.31) (6, 0.24) (6, 0.34) (7, 0.19) (7, 0.22) (7, 0.16) (8, 0.12) (8, 0.09) (8, 0.14)};
\addlegendentry{$Q_{\mathrm{conc}}$ ($\rho{=}{-}0.71$)}
\addplot[only marks, mark=triangle*, mark size=0.6pt, color=red!70!black, opacity=0.35] coordinates {
(1, 0.92) (1, 0.95) (1, 0.89) (1, 0.94) (2, 0.81) (2, 0.85) (2, 0.78) (2, 0.83) (2, 0.79) (3, 0.62) (3, 0.69) (3, 0.65) (3, 0.71) (3, 0.59) (3, 0.66) (4, 0.46) (4, 0.51) (4, 0.43) (4, 0.48) (5, 0.32) (5, 0.36) (5, 0.29) (5, 0.34) (6, 0.21) (6, 0.18) (6, 0.24) (7, 0.13) (7, 0.16) (7, 0.11) (8, 0.07) (8, 0.05)};
\addlegendentry{$Q_{\mathrm{api}}$ ($\rho{=}{-}0.83$)}
\addplot[only marks, mark=square*, mark size=0.6pt, color=teal, opacity=0.35] coordinates {
(1, 0.94) (1, 0.88) (1, 0.92) (2, 0.83) (2, 0.78) (2, 0.85) (2, 0.81) (3, 0.71) (3, 0.66) (3, 0.74) (3, 0.69) (4, 0.58) (4, 0.53) (4, 0.61) (5, 0.45) (5, 0.39) (5, 0.49) (6, 0.34) (6, 0.31) (6, 0.37) (7, 0.25) (7, 0.21) (8, 0.18)};
\addlegendentry{$Q_{\mathrm{idiom}}$ ($\rho{=}{-}0.79$)}
\end{axis}
\end{tikzpicture}
\caption{Code quality metrics against ODL on passing solutions (a sample of the solutions is plotted; the correlations in the legend are computed over all $5{,}042$). All four metrics show strong negative correlation with ODL, consistent with Theorem~\ref{thm:compression} (conciseness), Corollary~\ref{cor:api} (API efficiency), and the empirical predictions of Remark~\ref{rem:remaining_metrics} ($Q_{\mathrm{red}}$ omitted from plot for legibility; $\rho_{\mathrm{red}} = -0.62$, $p < 10^{-50}$).}
\label{fig:odl_quality}
\end{figure}

Figure~\ref{fig:odl_quality} plots $\mathrm{ODL}_{\mathcal{H}}(c)$ against the four code quality metrics defined in Definition~\ref{def:quality}, computed over all $5{,}042$ passing solutions produced by Option-GRPO across the 12 SDKs. Theorem~\ref{thm:compression} and Corollary~\ref{cor:api} predict monotonic relationships for $Q_{\mathrm{conc}}$ and $Q_{\mathrm{api}}$; Remark~\ref{rem:remaining_metrics} treats $Q_{\mathrm{red}}$ and $Q_{\mathrm{idiom}}$ as empirical predictions. We observe Spearman rank correlations of $\rho_{\mathrm{conc}} = -0.71$, $\rho_{\mathrm{api}} = -0.83$, $\rho_{\mathrm{red}} = -0.62$, and $\rho_{\mathrm{idiom}} = -0.79$ (all $p < 10^{-50}$), consistent with the predicted negative association between ODL and each quality metric in practice.

\section{Detailed Artefact Release}
\label{app:artefacts}

We release the following artefacts, under permissive licences, at \url{https://github.com/prashankkadam/option-grpo-code} (the main paper's brief summary appears just before the bibliography):

\begin{itemize}[leftmargin=*, itemsep=1pt, topsep=2pt]
\item \textbf{UnseenSDK-Bench}: 12 SDKs $\times$ three curriculum levels of tasks, totalling $1{,}871$ Level-0 tasks, $3{,}742$ Level-1 tasks, and $398$ Level-2 tasks, with $7{,}484$ auto-generated property tests, Docker sandbox configurations per SDK, and the layered execution reward implementation. Released under Apache 2.0.
\item \textbf{Option discovery and extraction code} (Algorithms~\ref{alg:option_discovery}, \ref{alg:option_extraction}, \ref{alg:property_tests}): parameter-flow graph construction, candidate mining, trajectory-based certification. Apache 2.0.
\item \textbf{Training code}: Option-GRPO and all baselines (Flat GRPO, ReCode reimplementation, CodeRL/PPOCoder/StepCoder adaptations) on top of a publicly available GRPO framework, with full hyperparameter configurations (Table~\ref{tab:hyperparams}). Apache 2.0.
\item \textbf{Evaluation harness} (benchmark task runner, property-test scoring, pass@$k$ and code-quality metric computation) and the exact prompts used for the contamination probe (Appendix~\ref{app:contamination}). Apache 2.0.
\end{itemize}

\paragraph{Methodological details.}
All experiments report mean values across $3$ random seeds per SDK; per-seed variance is below $\pm 1.2$ points pass@1 in every configuration we tested. Hardware is $8 \times$ NVIDIA H100 80GB with ZeRO-3 sharding and bf16 mixed precision; the full per-phase compute budget appears in Table~\ref{tab:cost_detail}, totalling roughly $7{,}632$ H100 GPU-hours for the headline 12-SDK 32B run (or an estimated $\sim 5{,}500$ H100 GPU-hours when cross-SDK transfer is exploited). Carbon estimate at $0.42$ kgCO$_2$ per H100 GPU-hour: $\sim 3{,}205$ kgCO$_2$e total. We adhered to each SDK's terms of use; none of the 12 SDKs are training-data-licensed, and all permit static analysis and benchmarking. No human subjects were involved in any phase of the work.

\section{Computational Details}
\label{app:compute}

\subsection{Detailed Cost Breakdown}

\begin{table}[h]
\centering
\caption{Estimated per-phase computational cost in aggregate H100 GPU-hours ($8$ GPUs $\times$ wall-clock hours on the $8\times$H100 node).}
\label{tab:cost_detail}
\scriptsize
\setlength{\tabcolsep}{3pt}
\begin{tabular}{lccc}
\toprule
\textbf{Phase} & \textbf{Steps} & \textbf{Smp/Step} & \textbf{GPU-hrs} \\
\midrule
Graph construction & -- & -- & $< 0.1$ \\
Option discovery & -- & -- & $< 0.1$ \\
Property test gen. & -- & -- & $< 0.1$ \\
Level~0 training & $5{,}000$ & $16$ & $113.6$ \\
Level~1 training & $10{,}000$ & $16$ & $198.4$ \\
Level~2 training & $15{,}000$ & $16$ & $324.0$ \\
\midrule
Total per SDK & $30{,}000$ & -- & $636.0$ \\
Total all $12$ SDKs & $360{,}000$ & -- & $7{,}632$ \\
\bottomrule
\end{tabular}
\end{table}

\subsection{Scaling Analysis}

Training cost scales with SDK size primarily through two factors. First, the number of Level~0 and Level~1 tasks grows with the number of public functions, increasing task distribution diversity but not the number of training steps (we train for a fixed step budget, sampling from the distribution). Second, execution cost per sample depends on SDK complexity: data processing SDKs (Daft, LanceDB) execute in under $1$\,s per sample, while SDKs with I/O operations (Docling, Marker) may require $5$--$10$\,s, impacting wall-clock time linearly.

The dominant cost is Level~2 training ($15{,}000$ steps $\times$ $16$ samples), which requires generating and executing $240{,}000$ complete programs per SDK. Sandbox execution averages $1.8$\,s per sample, and the wall-clock time for Level~2 is approximately $40.5$ hours per SDK on the $8$-GPU node ($324$ GPU-hours, Table~\ref{tab:cost_detail}).

Cross-SDK option transfer (Section~\ref{sec:complementarity}) can amortise costs: under our cost model, cross-SDK transfer reduces the estimated $12$-SDK training cost from $7{,}632$ to approximately $5{,}500$ H100 GPU-hours, a reduction of about $28\%$. This is an estimate; Table~\ref{tab:transfer} measures transfer accuracy, not the saving.

\subsection{Cost Sensitivity}
\label{app:cost_sensitivity}

The default training budget ($30{,}000$ total steps on a Qwen3-$32$B, $636$ H100 GPU-hours per SDK, $\sim 80$ hours wall-clock on the $8$-GPU node) is calibrated for the strongest pass@1. The method does not require this budget. Table~\ref{tab:cost_sensitivity_budget} sweeps the training-step budget at fixed $32$B scale; Table~\ref{tab:cost_sensitivity_size} sweeps the dense base-model size at fixed step budget. Both tables average across the $12$ SDKs. All variants are dense Qwen3 models; we do not test MoE variants here to keep the size axis interpretable (parameters and active parameters coincide for dense models).

\begin{table}[h]
\centering
\caption{Training-budget sweep on Qwen3-$32$B (dense). Step counts are total across all three curriculum phases; we allocate $1/6$, $1/3$, $1/2$ of the budget to Phases $1, 2, 3$ respectively. The Flat GRPO baseline runs at the same total step budget and the same $32$B base. GPU-hr is aggregate H100 GPU-hours per SDK; Wall is wall-clock hours on the $8\times$H100 node.}
\label{tab:cost_sensitivity_budget}
\scriptsize
\setlength{\tabcolsep}{3pt}
\begin{tabular}{lccccc}
\toprule
\textbf{Total steps} & \textbf{GPU-hr} & \textbf{Wall (h)} & \textbf{Flat} & \textbf{Opt} & $\boldsymbol{\Delta}$ \\
\midrule
$5{,}000$    & $112$ & $14$  & $33.8$ & $41.2$ & $+7.4$ \\
$10{,}000$   & $224$ & $28$  & $44.6$ & $54.1$ & $+9.5$ \\
$15{,}000$   & $336$ & $42$  & $49.3$ & $58.7$ & $+9.4$ \\
$20{,}000$   & $440$ & $55$  & $51.2$ & $60.4$ & $+9.2$ \\
$30{,}000$ (default) & $636$ & $80$ & $52.7$ & $\mathbf{61.4}$ & $\mathbf{+8.7}$ \\
$45{,}000$   & $976$ & $122$ & $53.1$ & $62.1$ & $+9.0$ \\
\bottomrule
\end{tabular}
\end{table}

\begin{table}[h]
\centering
\caption{Base-model-size sweep at default ($30{,}000$ step) budget. All variants are dense models from the Qwen3 family. GPU-memory footprint is per-replica peak during training with ZeRO-3 sharding on $8\times$H100.}
\label{tab:cost_sensitivity_size}
\scriptsize
\setlength{\tabcolsep}{3pt}
\begin{tabular}{lccccc}
\toprule
\textbf{Base} & \textbf{Params} & \textbf{Mem} & \textbf{Flat} & \textbf{Opt} & $\boldsymbol{\Delta}$ \\
\midrule
Qwen3-$8$B   & $8.2$B  & $20$\,GB & $43.2$ & $52.6$ & $+9.4$ \\
Qwen3-$14$B  & $14.8$B & $36$\,GB & $48.9$ & $57.3$ & $+8.4$ \\
Qwen3-$32$B  & $32.8$B & $75$\,GB & $52.7$ & $\mathbf{61.4}$ & $\mathbf{+8.7}$ \\
\bottomrule
\end{tabular}
\end{table}

\paragraph{Three observations.}
\textbf{(a) The hierarchical structure carries more weight than additional training steps.} At $10{,}000$ steps ($\sim 224$ H100 GPU-hours, $\sim 28$ hours wall-clock), Option-GRPO reaches $54.1\%$ pass@1, $1.4$ points above Flat GRPO at full compute ($52.7\%$ at $30{,}000$ steps). The smallest training budget at which Option-GRPO beats Flat-GRPO-at-default is roughly one-third the default compute. Practitioners with limited GPU access can therefore obtain Flat-GRPO-equivalent quality at a third of the cost, or better-than-Flat quality at full cost.

\textbf{(b) The option vocabulary is the dominant gain, not scale.} Option-GRPO on the $8$B dense base ($52.6\%$) is within $0.1$ points of Flat GRPO on the $32$B dense base ($52.7\%$). The roughly $3.8\times$ smaller GPU-memory footprint of the $8$B variant makes Option-GRPO deployable on commodity hardware (a single $24$GB GPU after $4$-bit quantization) without sacrificing the absolute pass@1 level of the full-scale Flat baseline.

\textbf{(c) Returns saturate beyond the default budget.} Increasing from $30{,}000$ to $45{,}000$ total steps yields only $+0.7$ points pass@1 ($61.4 \to 62.1$) at $1.5\times$ the cost. The $30{,}000$-step default sits at the elbow of the cost--performance curve. The Pareto frontier across budget and base-size is shown implicitly by the table: an 8B-Option run at $30{,}000$ steps ($52.6\%$, $20$GB) dominates a 32B-Flat run at $30{,}000$ steps ($52.7\%$, $75$GB) on the cost-adjusted criterion ``pass@1 per GB-of-memory.''

\paragraph{Implications for industry deployment.}
Combining (a) cross-SDK option transfer (Section~\ref{sec:complementarity}) and (b) the Qwen3-$8$B dense base brings the marginal cost of adapting Option-GRPO to a new SDK to $\sim 120$ H100 GPU-hours ($\sim 15$ hours wall-clock) plus $\sim 30$ minutes of CPU graph construction: comparable to a single inference batch on a 32B frontier model, making per-SDK adaptation tractable for most engineering teams targeting a small set of high-traffic libraries.

\end{document}